\documentclass[]{interact}

\usepackage{epstopdf}
\usepackage[caption=false]{subfig}

\usepackage[doublespacing]{setspace}

\usepackage[numbers,sort&compress,merge]{natbib}
\bibpunct[, ]{[}{]}{,}{n}{,}{,}
\renewcommand\bibfont{\fontsize{10}{12}\selectfont}

\theoremstyle{plain}

\theoremstyle{definition}

\theoremstyle{remark}

\usepackage{fp} 
\usepackage{bm} 
\usepackage{amssymb} 

\newcommand*{\vH}{\boldsymbol{H}}
\newcommand*{\vS}{\boldsymbol{S}}

\renewcommand*{\Pi}{{\varPi}}

\newcommand*{\vc}{\boldsymbol{c}}
\newcommand*{\vz}{\boldsymbol{z}}

\newcommand*{\vD}{\boldsymbol{D}}
\newcommand*{\vO}{\boldsymbol{O}}
\newcommand*{\vC}{\boldsymbol{C}}

\newcommand*{\AB}{{AB}}

\newcommand*{\nrho}{\rho^{'}}

\begin{document}
\articletype{ARTICLE}

\title{Chemical bond analysis for the entire periodic table:
Energy Decomposition and Natural Orbitals for Chemical Valence in the Four-Component Relativistic Framework}

\author{
\name{Diego Sorbelli,\textsuperscript{a,b} Paola Belanzoni,\textsuperscript{a,c} Loriano Storchi,\textsuperscript{d} Olivia Bizzarri,\textsuperscript{c} Beatrice Bizzarri\textsuperscript,{a,c} Edoardo Mosconi\textsuperscript{c} and Leonardo Belpassi\textsuperscript{c} \thanks{CONTACT L.~B. and L. S. Email: leonardo.belpassi@cnr.it;loriano@storchi.org}}
\affil{\textsuperscript{a}Dipartimento di Chimica, Biologia e Biotecnologie, Universit\`a degli Studi di Perugia,
Via Elce di Sotto 8, 06123
Perugia, Italia; \textsuperscript{b}Present address: Pritzker School of Molecular Engineering, University of Chicago, 5640 South Ellis Avenue, Chicago, IL, 5747 USA; \textsuperscript{c}Istituto di Scienze e Tecnologie Chimiche ``Giulio Natta'' (SCITEC-CNR), Consiglio Nazionale delle Ricerche
 c/o
Dipartimento di Chimica, Biologia e Biotecnologie, Universit\`a degli Studi di Perugia,
Via Elce di Sotto 8, 06123
Perugia, Italia; \textsuperscript{d} Dipartimento di Farmacia, Universit\`a degli Studi `G. D'Annunzio',
Via dei Vestini 31, 66100
Chieti, Italia}
}

\maketitle

\begin{abstract}
Chemical bonding is a ubiquitous concept in chemistry and it provides a
common basis for experimental and theoretical chemists to explain and
predict the structure, stability and reactivity of chemical species.
Among others, the Energy Decomposition Analysis (EDA, also known as
the Extended Transition State method)
in combination with Natural Orbitals for
Chemical Valence (EDA-NOCV)
is a very powerful tool for the analysis of the chemical  bonds
based on a charge and energy decomposition scheme within a common theoretical
framework.
While the approach has been  applied
in a variety of chemical contexts,
the current implementations of the EDA-NOCV scheme include
relativistic effects only at scalar level,
so simply neglecting the spin-orbit coupling effects and de facto
limiting its applicability.  In this work, we extend the EDA-NOCV
method to the relativistic four-component Dirac-Kohn-Sham theory that
variationally accounts for spin-orbit coupling. Its correctness
and numerical stability have been demonstrated
in the case of simple molecular systems, where the relativistic effects
play a negligible role, by comparison with the implementation
available in the ADF modelling suite (using the non-relativistic Hamiltonian and
the scalar ZORA approximation). As an illustrative example
we analyse the metal-ethylene coordination bond in
the group 6-element series (CO)$_5$TM-C$_2$H$_4$, with TM =Cr, Mo,
W, Sg, where relativistic effects are likely to play an increasingly
important role as one moves down the group.
The method provides a clear measure (also in
combination with the CD analysis) of the donation and back-donation
components in coordination bonds, even when relativistic effects,
including spin-orbit coupling, are crucial for understanding the chemical
bond involving heavy and superheavy atoms.

\resizebox{25pc}{!}{\includegraphics{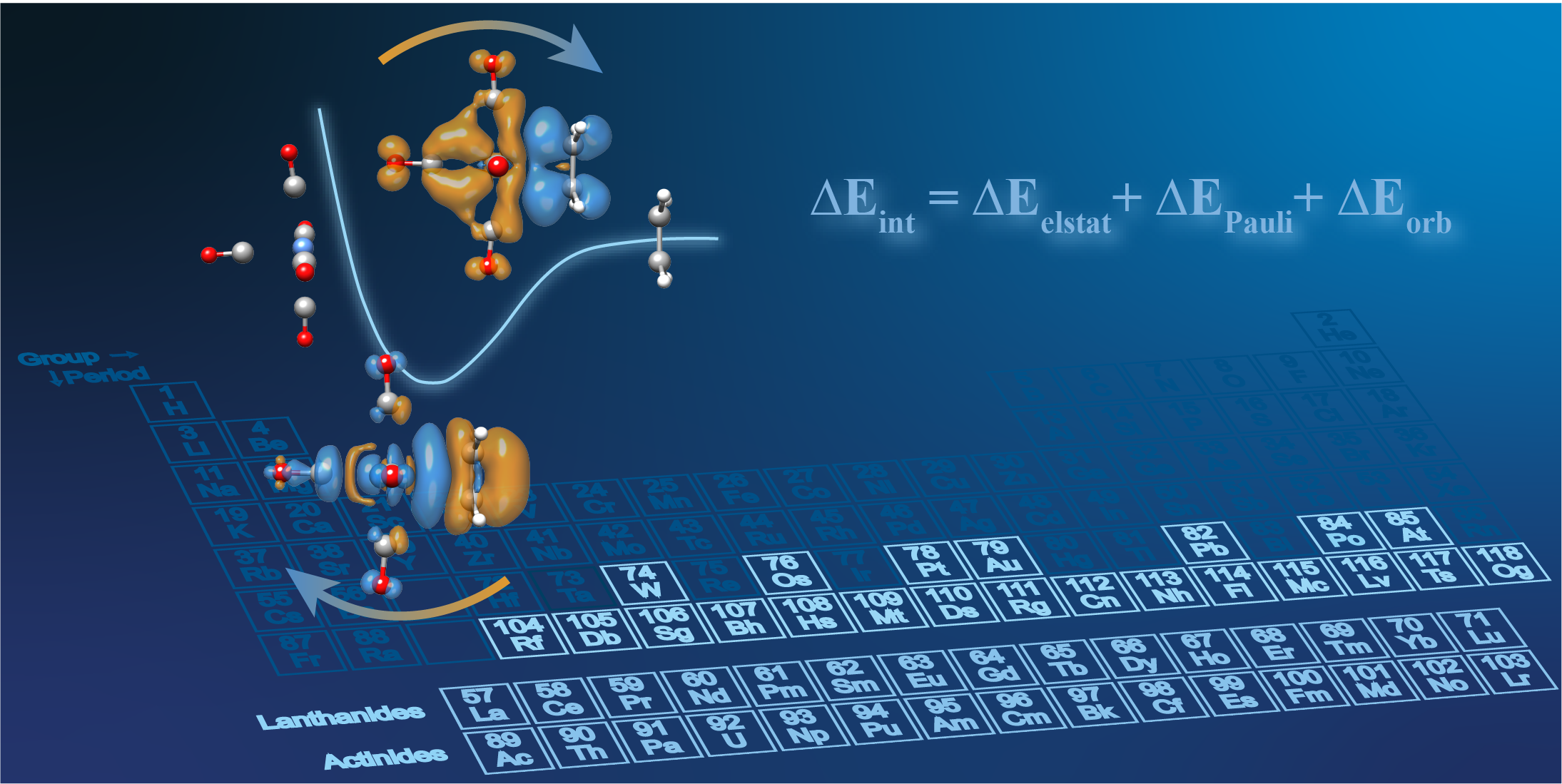}}
\end{abstract}


\section{Introduction}
In recent decades, several approaches, mainly in theoretical
chemistry, have been introduced to analyse and characterise chemical
bonding. These can be divided into two main groups: i) methods
mainly focusing on the decomposition of the binding energy into 'chemically
meaningful' contributions\cite{blwfed, kitaura, sapt, zieglerETS:1977,jacobsen}
and ii) methods that mainly focus on wave function or electron density
analysis\cite{bader67_3341,bader67_3381,bader2,esterhuysen,coppens,Schwarz1985,becke,mulli,Dapprich,Frenkingand,belpassi08_1048}.
As a matter of fact, a chemical bond is not a well-defined quantum mechanical
observable \cite{schwerdtfeger14_383} and so it is not surprising that
discussions about the nature of chemical bonds typically lead to disputes
and misunderstandings. So when designing, developing or applying a
theoretical method for analysing a chemical bond, it can be useful to
recall Coulson's perspective\cite{coulson:1952} which he expressed in a lecture in 1952:
``{\it Sometimes it seems to me that a bond between two atoms has become
so real, so tangible, so friendly, that I can almost see it. And then
I awake with a little shock: for a chemical bond is not a real thing:
it does not exist: no one has ever seen it, no- one ever can. It is a
figment of our own imagination...}'' and also ``{\it it is
a most convenient fiction, which, as we have seen, is convenient both
to experimental and theoretical chemists}''. If the fundamental theory
of quantum mechanics does not help to design unambiguous definitions for
chemical bonds, we believe that an important criterion for assessing the
validity of a model for chemical bonds lies in its predictive ability
within the chemical space.~\cite{schwerdtfeger14_383} For a
critical analysis of the most common bonding models, and the quantum
mechanical description of chemical bonds, we refer the reader to the
comprehensive review by Frenking et al.~\cite{Frenking:2019}, which also
includes a description of the main theoretical methods available today
for bonding analysis. For an overview of the physical nature of chemical
bonding, including the latest computational approaches, we refer the reader to
a very recent collection of articles~\cite{chembondJCP:2023}.

Among other methods available in the literature, energy decomposition analysis (EDA),
originally developed by Morokuma \cite{kitaura} and by Ziegler and
Rauk \cite{zieglerETS:1977} and also known as the Extended Transition
State Method (ETS), combined \cite{mitoraj2009,michalak2008bond}
with the original density partitioning method Natural Orbitals for the Chemical Valence (NOCV) of Mitoraj and
Michalak \cite{mitoraj07_347,mitoraj2007donor} represents a very powerful
method that bridges the gap between elementary quantum mechanics and a
conceptually simple interpretation of the nature of chemical bonds. The
method, known as EDA-NOCV (or ETS-NOCV), has become an important tool for the
analysis of bonds. Indeed, it takes into account different types of
physical interactions that contribute to the experimentally observable
bond dissociation energy. In particular, this approach is able to decompose
the orbital relaxation energy into NOCV pairwise contributions that can be
associated with a particular electronic deformation density. The latter can
be visualised to characterise the interaction. Furthermore, the NOCV
distribution can often be correlated with simple chemical concepts,
such as for instance the Dewar-Chatt-Duncanson (DCD) bonding model for coordination
chemistry. For a detailed and critical overview of the  EDA-NOCV method,
and its most common applications, we refer the interested reader to
Ref.~\cite{FrenkingWIREs:2018}

Some of us have shown that the rearrangement of electron
density associated with the NOCV orbitals pair can be analysed
quantitatively by using the so-called charge-displacement (CD)
analysis\cite{belpassi08_1048}. The method, also known as
NOCV/CD,~\cite{bistoni15_084112,doi:10.1021/acs.jctc.5b01166} is based
on the partial progressive integration of the rearrangement of
electron density that occurs during bond formation and is associated
with the NOCV partitioning of orbital relaxation.  The resulting
approach has been applied to quantify the donation and back-donation
components of the DCD model, which can be disentangled and
brought into very close correlation with experimental observations
\cite{bistoni13_11599, ciancaleoni2015selectively, C5DT02183A,
Bistoni:2016}. The method has been also applied to molecular systems without
symmetry constraints,~\cite{bistoni15_084112,marchione201713c,
CHEM:CHEM201700638,C7CP00982H} including the analysis of the
evolution of a chemical bond between catalyst and substrate along
the steps of a reaction pathway~\cite{bistoni15_084112}. More
recently, the  EDA-NOCV/CD method has helped us to
elucidate the peculiar nature of bonding in coinage metal-aluminyl
complexes\cite{jacs:2021,sorbelli_inorg:2022,sorbellirossi_inorg:2022}
and their reactivity with carbon dioxide and other small
molecules~\cite{sbb_inorchem:2022,chemscience:2023,https://doi.org/10.1002/chem.202203584}.

Despite its wide diffusion,  the current implementations of the EDA-NOCV method are limited to
non-relativistic Hamiltonians or to include relativistic effects  at scalar
level, typically using the scalar ZORA approximation,
which simply overlooks the spin-orbit coupling effects. However, it
is widely recognised that spin-orbit coupling can play an
important role not only in spectroscopy, but also in chemical
bonding~\cite{li1995relativistic,schwerdtfeger14_383,C1CP20512A}
and chemical
reactivity.~\cite{pyykko1988relativistic,gorin2007relativistic,schwarz2003relativistic,schwerdtfeger1992low,C6SC02161A,ANIE:ANIE201606001,Ricciarelli:2020,Gaggioli:2018}
The development of methods which are able to analyse the chemical bonds
within a theoretical framework which incorporates spin-orbit coupling
effects is of high importance. Among
others,~\cite{Saue:2001,Reiher:2007,Sablon2010,Klobukowski:2011,Galland:2014}
the recent work of Senjen et al.~\cite{Senjean:2021} where
the intrinsic atomic and bond orbitals (IBOs) scheme has been generalised
to fully relativistic applications using complex and quaternion spinors as implemented in
the DIRAC code~\cite{DIRAC:2020,DIRAC23} goes exactly in this direction. Some of us have extended
the NOCV scheme to the Dirac-Kohn-Sham module of the code BERTHA
(and in its new Python API, PyBERTHA) and
the approach has been successfully used (also in combination with the CD analysis,
NOCV/CD) to study chemical bonding and s-d hybridisation in Group 11 M-CN
cyanides (M = Cu, Ag and Au) and in Group 11 MH$_2^-$ dihydrides (M=Cu, Ag, Au, Rg)\cite{doi:10.1021/acs.jpca.0c09043}, superheavy elements interacting with metal
clusters, and to characterise weakly bound systems involving astatine \cite{de2018charge,rossi2020spin,de2019chemical}.
Despite the quantitative measure of the charge transfer (CT) involved upon a bond formation, the NOCV/CD approach,
is exclusively focused on the analysis of the rearrangement of the electron density
and gives no indication of the different energy
terms that contribute to the total energy of the bond.  It therefore seems
appropriate to extend the NOCV approach already implemented in BERTHA \cite{de2018charge} and combine
it with EDA. This combined scheme offers  the possibility to obtain quantitative
information about different energetic terms contributing to the total
interaction energy and to assign a well-defined energetic contribution
to the rearrangement of the electron density associated with a specific
pair of NOCV orbitals.
In this paper we present the formalism and implementation
of the EDA-NOCV scheme in the context of the relativistic
four-component DKS method, where the spin-orbit coupling is
variationally included. In Section \ref{methodsec} we recall
the essential aspects of the EDA method, including the Extended Transition
State method that is used in combination with the NOCV theory
(EDA-NOCV). We describe how the scheme can be extended
to the Dirac-Kohn-Sham theory as is currently implemented in the DKS module
of the BERTHA
code\cite{belpassi11_12368,storchi10_384,storchi13_5356,rampino14_3766}
taking advantage of the new Python API. In section \ref{sec:results} we
present first test calculations to validate the correctness and numerical
stability of our new implementation. An application is also reported
in order to illustrate the effective usefulness of the approach to analise the coordinative bond
in a consistent manner in the whole periodic table.
In particular we have investigated the metal-ethylene bond in the full set
of the group 6  carbonyl complexes (CO)$_5$TM-C$_2$H$_4$, with TM =Cr, Mo, W and Sg (Seaborgium, element 106),
where relativistic effects become increasingly important. Some conclusions and
perspectives are drawn in the last Section.

\section{Methodology}\label{methodsec}

\subsection{Energy Decomposition Analysis and Natural Orbital for the
Chemical Valence (EDA-NOCV)}

We begin this Section with a brief overview of the EDA scheme\cite{}
in the version originally introduced by Morokuma \cite{kitaura} and
successively  by Ziegler and Rauk \cite{zieglerETS:1977} at the Hartree-Fock or
Hartree-Fock-Slater level. In the literature, this scheme is also known
as the Extended Transition State (ETS) scheme, named after the original work
of Ziegler and Rauk \cite{zieglerETS:1977}.
An important impetus for the dissemination of EDA was its development in the
context of Kohn-Sham theory and its
implementation in the ADF code over many years~\cite{ch1BB:2000}. More
recently, the scheme has been combined with natural orbitals for
chemical valence (EDA-NOCV)~\cite{mitoraj09_962},
where the Extended Transition State method was used to decompose  the orbital
interaction energy in terms of well defined contributions associated
with specific NOCV deformation densities that can be visualized
and  analised to characterize the interaction.
For a detailed description, we refer the reader to the
original papers~\cite{zieglerETS:1977,ch1BB:2000,mitoraj09_962}. For
the sake of clarity, in the following summary of the EDA-NOCV scheme,
we attempt to use a notation which is as consistent as possible with that used
by Frenking et al. in a recent review
article \cite{FrenkingWIREs:2018} and by Ziegler et al.
in the seminal work on ETS-NOCV \cite{mitoraj09_962}.

The fundamental idea of this energy partition scheme is the decomposition of the total binding energy
($\Delta E$) between two fragments A and B as a sum of well-defined terms.
These terms can be defined
by assigning intermediate states to the system in the course of
bond formation in a kind of stepwise mechanism.
Each state can be reached by applying  a well-defined mathematical procedure
that can be associated with a {\it physically} meaningful entity.

At first the total binding energy ($\Delta E$) is split into two main components, $\Delta E_{prep}$
and $\Delta E_{int}$:
\begin{equation}\label{eq:totalenergy} \Delta E =
\Delta E_{prep} + \Delta E_{int}
\end{equation}
$\Delta E_{prep}$ is
the energy required to bring the
 fragments A and B from their equilibrium geometry, at their ground electronic configuration,
to the geometry they acquire
in the compound AB in their valence electronic configuration. Instead, $\Delta E_{int}$ is the instantaneous interaction
energy between the two fragments in the molecule. The latter quantity
is in turn divided into three terms:
\begin{equation}\label{eq:EDA} \Delta E_{int} = \Delta E_{elstat} +
\Delta E_{Pauli} + \Delta E_{orb}
\end{equation}
where $\Delta E_{elstat}$
is the electrostatic interaction, $\Delta E_{Pauli}$ is called the Pauli
term and $\Delta E_{orb}$ is the orbital interaction term.
\begin{figure}[htb]
\includegraphics[width=1.\textwidth]{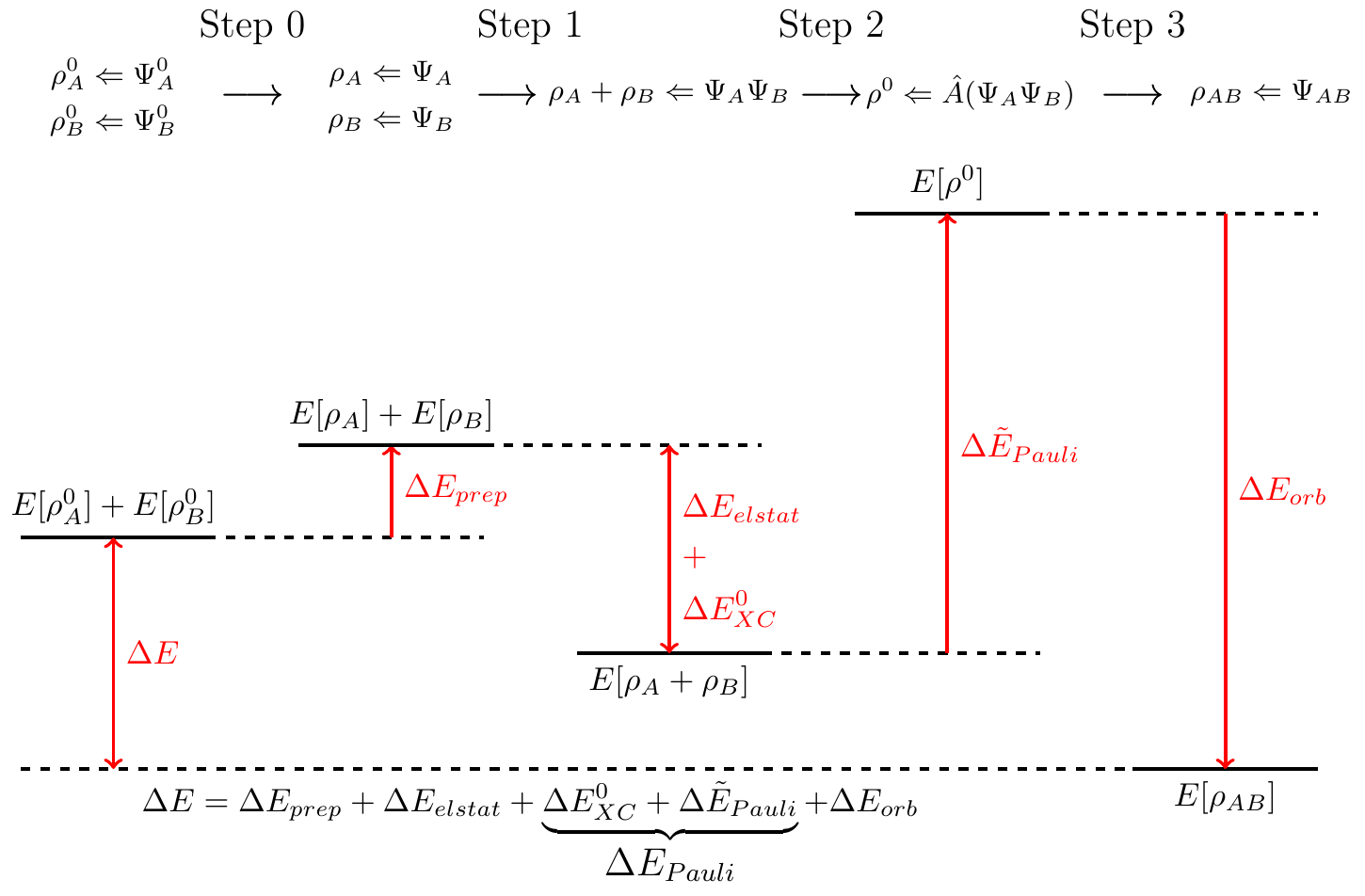}
\caption{Graphycal stepwise representation of EDA, see text for details.}\label{edascheme}
\end{figure}

The graphical illustration of the individual steps at the basis of EDA can be
visualized in Figure~\ref{edascheme}.
Initially,
the system switches from state 0, which is characterised by the isolated and non-interacting fragments (infinite spatial distance)
with well-defined Kohn-Sham determinants ($\Psi_A^0$ and $\Psi_B^0$) and electron densities
($\rho_A^0$ and $\rho_B^0$). In the first step, the fragments are
distorted into the geometry ($\Psi_A$ and $\Psi_B$) they possess in the final adduct.
The energy increases by an amount that corresponds to the definition of the preparation energy mentioned above, $\Delta
E_{prep}=E[\rho_A]+E[\rho_B]-E[\rho_A^0]-E[\rho_B^0]$. In the second step, the distorted fragments are moved from infinite distance to their final position, which they occupy in the complex, without changing their densities
($\rho_A$ and $\rho_B$) and orbitals. The associated change in energy is now given
by $E[\rho_A+\rho_B]-E[\rho_A]-E[\rho_B]$
which can be easily written  as (see Appendix \ref{appendix:a})
\begin{equation}\label{eq:secondstep}
E[\rho_A+\rho_B]-E[\rho_A]-E[\rho_B] = \Delta E_{elstat} + \Delta E_{XC}^0.
\end{equation}
$\Delta E_{elstat}$ represents the classical electrostatic interaction between the unperturbed charge distributions ($\rho_A$ and $\rho_B$)
of the prepared fragments
when they are brought together at their final positions. The resulting total
density is a simple superposition of the fragment densities, and
its explicit expression is given by
\begin{eqnarray*}
\Delta E_{elstat} & = & -\sum_{i \in \{A\}} \int \frac{Z_i \rho_B(r)}{|R_i-r|} dr -\sum_{i \in \{B\}} \int \frac{Z_i \rho_A(r)}{|R_i-r|} dr + \\
                    & + & \sum_{\substack{i \in \{A\} \\j \in \{B\}}} \frac{Z_i Z_j}{|R_j-R_i|} + \int \int \frac{(\rho_A(r_1)\rho_B(r_2))}{|r_1-r_2|} dr_1 dr_2,
\end{eqnarray*}
where the first two terms represent the Coulomb interaction between the charges of the nuclei of fragment
A (and B) and the one-electron density of fragment B (and A). Instead, the third term is the classical electrostatic repulsion between  the
nuclei. Finally, the last term is the Coulomb repulsion between the one-electron densities associated with the two isolated
fragments ($\rho_A$ and $\rho_B$).
For the neutral fragments, $\Delta E_{elstat}$ is typically attractive because of a charge penetration that may occur between the two fragments.
The term $\Delta E_{ XC }^0$ in Eq.\ref{eq:secondstep}, defined as $E_{ XC }[\rho_A+\rho_B] - E_{ XC }[\rho_A] - E_{ XC }[\rho_B]$,
represents the corresponding change in the Kohn-Sham exchange correlation energy.
The electrostatic energy enters the first term of the EDA, see Eq.~\ref{eq:EDA}.
In this second step, the total energy $E[\rho_A+\rho_B]$ and the total electron density ($\rho_A+\rho_B$)
are associated with a wave function that is the simple product ($\Psi_A\Psi_B$) of the non-interacting Kohn-Sham determinants of fragments A and B. This product, of course, does not have the correct asymmetry property required by quantum mechanics for a fermionic system.

In the third step, an energy change occurs due to the transition from $\Psi_A\Psi_B$ to the wave function
$\Psi^0 = N\hat{A}[\Psi_A\Psi_B]$ (with an associated electron density $\rho^0$), which obeys the Pauli principle by
explicit antisymmetrisation ($\hat{A}$ operator) and renormalisation
(N constant) of the product of the fragments wavefunctions.
This is typically carried out via a simple orthonormalisation procedure of the occupied
orbitals of the fragments.
This step is associated with an energy increase of the system ($\Delta \tilde
E_{Pauli}$) that is defined as:
\begin{equation}
        \Delta \tilde E_{Pauli}=E[\rho^0]-E[\rho_A+\rho_B]
\end{equation}

It is important to note that the term $\Delta \tilde E_{Pauli}$ we have just
defined is not exactly  the same that appears in the EDA partitioning of Eq.\ref{eq:EDA}.
In fact, it is common in the literature to sum
$\Delta \tilde E_{Pauli}$ and $\Delta E_{ XC }^0$ to obtain the total Pauli or "exchange repulsion" term $\Delta E_{Pauli}$
\begin{equation}
\Delta E_{Pauli} = \Delta \tilde E_{Pauli} + \Delta E_{XC}^0.
\end{equation}
The reason for  such choice seems
to be related\cite{mitoraj09_962} with the fact that the positive and destabilising term $\Delta \tilde E_{Pauli}$
is dominant over $\Delta E_{XC}^0$.
This definition is used for instance in the EDA implementation of the ADF program.
For the present authors, however, this choice contains a certain degree of arbitrariness,
so, for the sake of clarity, we will report both   the entire Pauli term ($\Delta E_{Pauli}$) and also the two constituting terms
 ($\tilde \Delta E_{Pauli}$ and $\Delta E_{XC}^0$), separately.

Finally, the last term of Eq.~\ref{eq:EDA} is called the orbital
interaction energy $\Delta E_{orb}$. It is calculated in the fourth (and
last) step of the EDA scheme.  Here, one allows
$\Psi^0$ to relax to the fully converged Kohn-Sham determinant $\Psi_{
AB }$ (the associated density is denoted as $\rho$).  $\Delta E_{orb}$ accounts for the chemical contribution
of the interaction including electron pair,
charge transfer (e.g. HOMO-LUMO interactions) and polarisation (mixing
of empty/occupied orbitals on one fragment due to the presence of
another fragment). It is evident that polarisation and charge transfer
energy stabilization between fragments cannot be separated and both contribute to the $\Delta E_{orb}$.

All energy contributions in the EDA partitioning scheme have well-defined mathematical
definitions, but we have to recognise that none of them is an observable,
although they sum
to the experimentally measurable bond dissociation
energy (see Eq.~\ref{eq:totalenergy} and Eq.~\ref{eq:EDA}). Furthermore, the attentive reader
will no doubt have noticed a theoretical issue
when we use EDA within the Kohn-Sham framework. In particular, in step 2 of the
EDA scheme, the system has an energy that has been defined as a functional
of a density given by the sum of the density of the fragments  (this sum
appears in $E[\rho_A+\rho_B]$ and also in $\Delta E^0_{ XC }=E_{ XC
}[\rho_A+\rho_B]-E_{ XC }[\rho_A]-E_{ XC }[\rho_B]$)
which is not N-representable. Thus,  it is not clear which exchange-correlation contribution is describing. This has
already been discussed in details by Bickelhaupt and Baerends
at pag. 11 of Ref.\cite{ch1BB:2000}. The basic conclusion is that the exchange-correlation contribution is typically small with respect to
the other terms of the interaction (kinetic and potential energy), thus one
may be confident that  $E^0_{ XC}$ gives a reasonable representation
of the exchange-correlation energy also in this intermediate step. Therefor,
its use has been pragmatically accepted in the literature.
Based on the transition state method, Ziegler and Rauk showed, in their seminal paper
on ETS~\cite{zieglerETS:1977}, that one can write
 approximated expressions for the energy difference associated with a well-defined
rearrangement of the electron density.  In particular, it can be
shown that for the orbital interaction $\Delta E_{orb}$ the following expression is valid
\begin{eqnarray*}
    \Delta E_{orb} & =  & E[\rho]-E[\rho^0] \\
                        & = & \sum_{\mu,\nu} \Delta D_{\mu \nu}
                        F_{\mu\nu}[\rho_{orb}^T]+O(\Delta {\bf D}^3)
\end{eqnarray*}
where $\Delta D_{\mu \nu}$ are the matrix elements of
the  $\Delta {\bf D} = {\bf D^{ AB } -D^{0}}$ matrix, defined as the
difference of the density matrices associated with the full relaxed
density of the adduct
($\rho(r) = \sum_{\mu,\nu} \Delta D_{\mu \nu}^{
AB } \chi^*_{\mu}(r) \chi_{\nu}(r)$, step 4) and with the orthonormalized
density matrix ($\rho^0(r) = \sum_{\mu,\nu} \Delta D^0_{\mu ,\nu}
\chi^{*}_{\mu}(r) \chi_{\nu}(r)$, step 3).
$F_{\mu\nu}[\rho_{orb}^T]$ are the matrix elements of the
Kohn-Sham operator associated with the transition state density,
$\rho_{orb}^T=1/2(\rho^0(r) + \rho_{ AB }(r))$.
 The
basic idea is to expand the energy of both the final state ($E[\rho_f]$)
and the initial state ($E[\rho_i]$) in a Taylor series with respect to
a common origin, $E[\rho^T]$, which is the energy associated with the
average density ($\rho^T = 1/2(\rho_f +\rho_i)$).  For the reader
convenience, we give the derivation of the formula for the orbital
interaction energy corrected to O($\Delta {\bf D}^2$) and also the
formulae derived by Ziegler and corrected to O($\Delta {\bf D}^4$) (see
Appendix \ref{appendix:b}).  The use of the extended transition state
density in combination with the NOCV method was proposed by Ziegler et
al. in 2009\cite{mitoraj09_962}
and represents the key idea of the EDA-NOCV method.
This method allows for the splitting of the orbital interaction
energy into pairwise contributions based on the NOCV partitioning of
the associated electron density rearrangement (step 4).
It provides a consistent energy/charge partitioning
scheme where the different chemical contributions to the orbital relaxation can
be easily identified by visualising the associated deformation density
and quantified by providing the corresponding energy contributions.

In the NOCV scheme the
deformation density associated with the orbital relaxation in step 4,
can be brought into a diagonal form
in terms of NOCVs:
\begin{equation}
        \Delta \nrho (r)= \rho(r) - \rho^0 (r)= \sum_k v_k
        \left(
                |\phi_{k}|^2 - |\phi_{-k}|^2
        \right) =
        \sum_k \Delta \nrho_k(r)
\end{equation}
where $k$, a positive integer, numbers the NOCV pairs in descending order of $v_k$ ($v_k > 0$).

The NOCVs orbitals, $\phi_{\pm k}$, are defined as the eigenfunctions  of the so-called
'valence operator' of the valence theory of Nalewajski and Mrozek
\cite{nalewajski93_463,nalewajski94_187,nalewajski97_589},
which, with respect to the occupied spin orbitals of the molecule ($\phi_i^\AB$) and of the promolecule ($\phi_i^0$), is defined as
\begin{equation}\label{eq:voperator}
\hat{V} = \sum_i \left( | \phi_i^\AB \rangle \langle \phi_i^\AB | - | \phi_i^0 \rangle \langle \phi_i^0 | \right)
\; .
\end{equation}
The NOCVs have the special property that they can be grouped into pairs of complementary orbitals ($\phi_{k}, \phi_{-k}$) corresponding to eigenvalues with the same absolute value but opposite sign
(for the algebraic properties of NOCVs, see Ref. \cite{radon08_337}).
The spectral representation of $\hat{V}$ is given by
\begin{equation}\label{eq:Vspectral}
\hat{V} = \sum_k \pm v_k |\phi_{\pm k} \rangle \langle \phi_{\pm k} | 
\end{equation}

From the definition of the operator $\hat{V}$, it is clear that $k$ ranges from $1$ to the number of occupied spin orbitals.
Moreover, it is important to note that only a small subset of the
NOCV pairs correspond to values of $v_k$ that are significantly different from zero.
This means that only a few NOCVs are important for describing the rearrangement of electron density due to the bond formation.
Supposing that the basis set $\{\chi_i\}$ for the adduct (AB) is obtained by joining the basis sets of the two fragments (A and B),
this leads to a simple algebraic formulation
in which the chemical-valence operator can be written as follows
\begin{equation}
\hat{V} = \sum_{\mu \nu}
\Delta D_{\mu \nu} | \chi^*_{\mu} \rangle \langle \chi_{\nu} |
\end{equation}
where $\Delta D_{\mu \nu}$ are the elements of the density matrix
(${\bf \Delta D}= {\bf D} - {\bf D}^{0}$) defined above.
In this algebraic solution, the NOCV orbitals ($\phi_k$)
in Eq.~\ref{eq:Vspectral} are given by
\begin{equation} \phi_k = \sum_\mu z_{i,k} \chi_\mu
\end{equation}
where $z_{i,k}$ are obtained by the
solution of the generalised eigenvalue problem,
\begin{equation}
\bm{V}\bm{z}_k = v_k\,\bm{S}\,\bm{z}_k,
\end{equation}
where $\bm{V}$ is the matrix representation of the chemical-valence operator defined as
\begin{equation}
{\bf V}= {\bf S}\,{\bf \Delta D}\,{\bf S}
\end{equation}
and $\bf S$ is the overlap matrix.
The NOCVs can be constructed by diagonalising the matrix representation of the 'valence operator'
to ensure that the normalisation condition is satisfied.
It is interesting that comparing different representations of $\hat{V}$, we can easily derive a simple partition of the one-particle
deformation density matrix, ${\bm \Delta D}$,
associated with the NOCVs pairs ($\phi_{\pm k}$):
\begin{eqnarray*}
    \hat{V} & = & \sum_{\mu \nu} \Delta D_{\mu \nu} |\chi_{\mu} \rangle \langle \chi_{\nu}| \\ & = & \sum_k \pm v_k |\phi_{\pm k} \rangle \langle \phi_{\pm k} | \\ & = & \sum_k \sum_{\mu \nu} v_k (z^*_{\mu,k}z_{\nu,k} - z^*_{\mu,-k}z_{\nu,-k}) |\chi_{\mu} \rangle \langle \chi_{\nu}| \\ & = & \sum_{\mu \nu} (\sum_k \Delta D^k_{\mu \nu}) |\chi_{\mu} \rangle \langle \chi_{\nu}|
\end{eqnarray*}
where we define $\Delta D_{\mu,nu}^{(k)}= \nu_k (z^*_{\mu,k}z_{\nu,k} - z^*_{\mu,-k}z_{\nu,-k})$.
The sum of $\Delta D_{\mu,\nu}^{(k)}$ over $k$ clearly gives $\Delta D_{\mu \nu}$.

The rearrangement of the electron density associated with the k-th
NOCVs pair can be determined via the corresponding contribution to the
one-electron density matrix,  $\Delta D_{\mu,\nu}^{(k)}$
\begin{eqnarray}
\Delta \rho_k (r) & = &
v_k \left( |\phi_{k}|^2 - |\phi_{-k}|^2 \right) \\ & = &
v_k \sum_{\mu\nu} (z^*_{\mu,k}z_{\nu,k} - z^*_{\mu,-k}z_{\nu,-k})
\chi_{\mu}^*(r) \chi_\nu(r) \\ & = &  \sum_{\mu\nu} \Delta
D_{\mu \nu}^{(k)} \chi_{\mu}^*(r) \chi_\nu(r).
\end{eqnarray}

The partition of $\Delta D_{\mu \nu}$ into the contribution
associated with the NOCV pairs ($\Delta D_{\mu \nu}^{(k)}$) can be
directly applied to the partition of the total orbital interaction energy:
\begin{eqnarray*}
        \Delta E_{orb}  & =    & E[\rho]-E[\rho^0] \\
                        & = & \sum_{\mu,\nu} \Delta D_{\mu \nu} F_{\mu\nu}[\rho_{orb}^T] +O(\Delta D^3) \\
                        & = & \sum_k \sum_{\mu,\nu} \Delta D_{\mu \nu}^{(k)} F_{\mu\nu}[\rho_{orb}^T] +O(\Delta D^3)\\
                        & = & \sum_k E^{k} +O(\Delta D^3), \\
\end{eqnarray*}
where $E^{k}=\sum_{\mu,\nu} \Delta D_{\mu \nu}^{(k)} F_{\mu\nu}[\rho_{orb}^T]$.
The NOCV scheme applied in combination with  the transition state method
provides an unambiguous procedure to partition the
total orbital energy in contributions ($E^{k}$) associated with the specific charge rearrangements ($\Delta \rho_k (r)$).
We note that the error that arises when one uses the transition state method is very small,
or negligible, in most of the cases. An even more accurate formula (corrected up to $O(\Delta D^4)$)
has already been presented~\cite{zieglerETS:1977},
where $F_{\mu\nu}[\rho_{orb}^T]$ is replaced by
$2/3(F_{\mu\nu}[\rho_{orb}^T]+1/6F_{\mu\nu}[\rho^0]+1/6F_{\mu\nu}[\rho])$.
However, for most applications, this correction is usually negligible. For the sake of completeness, in Appendix B, we present
the basic derivations of the transition state formula and some numerical tests
for an illustrative molecular system considered in this paper have been presented in SI.

\subsection{Implementation of the EDA-NOCV scheme within Dirac-Kohn-Sham
module of BERTHA}

The EDA-NOCV method has been implemented within the Dirac-Kohn-Sham (DKS) module of the BERTHA code and in particular taking advantage of its
Python API, PyBERTHA.
The basic theory  of the DKS method and its implementation in BERTHA
has been described in detail in Refs.~\cite{belpassi11_12368,Belpassi2020}, including interesting details of our
memory open-ended implementation code development based on OpenMP~\cite{chandra2001parallel}
and the description of our Python API framework, PyBERTHA~\cite{Belpassi2020}.
Below we will briefly summarise the most important aspects of the DKS formalism and approximations as currently
implemented in BERTHA,
which are of some relevance for the implementation of the EDA-NOCV scheme.
In atomic units, and including only the longitudinal electrostatic potential, the DKS equation implemented in BERTHA is
\begin{equation}\label{eq:dks} \{c \bm{\alpha} \cdot {\bf p}+ {\beta}
c^2+v^{(l)}({\bm r})\} {\bm \Psi}_{i}({\bm r})=\varepsilon_i{\bm
\Psi}_{i}({\bm r}), \end{equation}
where $c$ is the speed of light in vacuum, $\bf p$ is the electron
momentum,
\begin{equation}
{\bm\alpha} = \left(
\begin{array}{cc}
0 & {\bm\sigma} \\
{\bm\sigma} & 0
\end{array}
\right)\ \mbox{and}\
{\beta} =
\left(
\begin{array}{cc}
I & 0 \\
0 & -I
\end{array}
\right)
\end{equation}

where ${\bm\sigma}=(\sigma_x,\sigma_y,\sigma_z)$, $\sigma_q$ is a
$2\times 2$ Pauli spin matrix and $I$
is a $2\times 2$ identity matrix.

A four-spinor solution of Eq.~(\ref{eq:dks}) is of the form
\begin{equation}
\Psi_i({\bm r})= \left[ \begin{array}{c}
                          \psi_i^{(1)}({\bm r}) \\
                          \psi_i^{(2)}({\bm r}) \\
                          \psi_i^{(3)}({\bm r}) \\
                          \psi_i^{(4)}({\bm r})
                        \end{array} \right]
\end{equation}
and the total relativistic charge-density is a scalar function, as in non-relativistic
context, and can be readily evaluated
as the scalar product of 4-component spinors according to
\begin{equation}\label{density}
        \rho({\bm r})=\sum_a {\Psi}_{a}({\bm r})^{\dagger}{\Psi}_{a}({\bm r})
\end{equation}
where the sum extends only over the occupied positive-energy
bound states (electronic states).
The longitudinal interaction term, $v^{(l)}({\bm r})$,  is represented by a diagonal operator borrowed from
non-relativistic theory and made up of a nuclear potential term $v_{\mathrm{N}}(\bm{r})$,
a Coulomb interaction term $v^{(l)}_{\mathrm{H}}[\rho(\bm{r})]$
and the exchange-correlation term $v^{(l)}_{\mathrm{XC}}[\rho(\bm{r})]$.
We mention that the Breit interaction contributes to the transverse part of the Hartree
interaction and it is not considered here.
In the present implementation of BERTHA we use exchange-correlation functionals and associated potentials depending only on the electron density, rather than on the relativistic four-current, and
pragmatically non-relativistic density functionals are used that were not explicitly
designed for use in relativistic calculations.
BERTHA is currently interfaced with the LIBXC library, which provides a portable,
well tested and reliable set of exchange and correlation functionals that are
used by several non-relativistic DFT codes.
LDA, GGA and meta-GGA type exchange-correlation functionals
can be used in what is called the "density-only" approximation\cite{jacob12_3661},
which means that the exchange-correlation functional depends only on the electron density,
its gradient (in the case of GGA) and not on other variables such as spin density or
magnetization\cite{jacob12_3661}, which may also be used to reparametrize the
exchange-correlation potential.

The spinor solution of Eq.~\ref{eq:dks} is expressed as a linear combination of  G-spinor basis functions
\cite{grant2007relativistic} ($M_\mu^T(\bm{r})$ with  $T=L,S$):

\begin{align}\label{eq:spin_sol}
\bm{\Psi}_i(\bm{r}) &= \begin{bmatrix}
                      \sum_{\mu = 1}^{N} c^{(L)}_{\mu i}M^{(L)}_{\mu}(\bm{r})  \\
                      \\
                      \sum_{\mu = 1}^{N} c^{(S)}_{\mu i}M^{(S)}_{\mu}(\bm{r})
                     \end{bmatrix}
\end{align}
where $L$ and $S$ refer to the so-called ``large'' and ``small'' component, respectively,
and the $c^{(T)}_{\mu i}$ are expansion coefficients to be determined. The collective index $\mu$ works as a tag
for the set of parameters (coordinates of the local origin and exponent, fine-structure quantum
number and magnetic quantum number) completely characterizing
the Gaussian-based two-component objects $M_\mu^T(\bm{r})$\cite{grant2007relativistic}.

The matrix representation of the DKS operator in the G-spinor basis is given by

\begin{align}\label{eq:H_DKS}
\bm{H}_{DKS} &= \begin{bmatrix}
                      \bm{V}^{(LL)} +mc^2 \bm{S}^{(LL)} & c\bm{\Pi}^{(LS)}  \\
                                   c\bm{\Pi}^{(SL)}      &  \bm{V}^{(SS)} -mc^2\bm{S}^{(SS)}
                     \end{bmatrix}
\end{align}

where $\bm{V}^{(TT)} = \bm{v}^{(TT)} + \bm{J}^{(TT)} + \bm{K}^{(TT)}$, with $T=L,S$.

The eigenvalue equation is

\begin{equation}
\bm{H}_{DKS}\begin{bmatrix} \bm{c}^{(L)} \\ 
                             \bm{c}^{(S)} 
                              \end{bmatrix} = E 
                              \begin{bmatrix} \bm{S}^{(LL)} & 0 \\
                                0 & \bm{S}^{(SS)} 
                                \end{bmatrix} 
                                \begin{bmatrix} 
                                   \bm{c}^{(L)}\\ 
                                   \bm{c}^{(S)} \end{bmatrix} 
\end{equation}

where $\bm{c}^{(T)}$ are the spinor expansion vectors of Eq. \ref{eq:spin_sol}. The $\bm{H}_{DKS}$ matrix is defined
in terms of $\bm{v}^{(TT)}, \bm{J}^{(TT)}, \bm{K}^{(TT)}, \bm{S}^{(TT)}$, and $\bm{\Pi}^{(TT')}$ matrices,
being respectively the basis representation of the nuclear, Coulomb, and exchange-correlation potential, the overlap matrix, and
the matrix of the kinetic operator.
The nuclear charges have been modeled by a finite
Gaussian distribution~\cite{Visscher:1997}.
The resulting matrix elements are defined by
\begin{align}
  & \bm{v}_{\mu \nu}^{(TT)} = \int \bm{v}_{N}(\bm{r})\rho_{\mu \nu}^{(TT)}(\bm{r})d\bm{r}   \\
  & J_{\mu \nu}^{(TT)} = \int v_{H}^{(\mathrm{l})}[\rho(\bm{r})]\rho_{\mu \nu}^{(TT)}(\bm{r})d\bm{r}\label{eq:ref1}   \\
  & K_{\mu \nu}^{(TT)} = \int v_{\mathrm{xc}}^{(\mathrm{l})}[\rho(\bm{r})]\rho_{\mu \nu}^{(TT)}(\bm{r})d\bm{r}\label{eq:ref2}  \\
  & S_{\mu \nu}^{(TT)} = \int \rho_{\mu \nu}^{(TT)}(\bm{r})d\bm{r} \label{ss} \\
  & \Pi_{\mu \nu}^{TT'} = \int M_\mu^{(T) \dagger}(\bm{r})(\bm{\sigma}\cdot \bm{p}) M_\nu  ^{(T')}(\bm{r})d\bm{r} .
\end{align}

The terms $\rho_{\mu\nu}^{(TT)}({\bm r})$ are the G-spinor overlap densities:
\begin{eqnarray}
\rho_{\mu\nu}^{(TT)}({\bm r}) & = &
M^{(T) \dagger}_{\mu}({\bm r}) M^{(T)}_{\nu}({\bm r})
\end{eqnarray}
These are  expressed as  finite superpositions,
of standard Hermite Gaussian-type
functions (HGTFs)
(see, e.g., Ref.~\citenum{saunders1983molecular}) which allows to use stardard
techniques for the analytical evaluation of the two-electron repulsion integrals.
The $\bm{H}_{DKS}$ matrix depends on $\rho(\bm{r})$
in $v_{\mathrm{xc}}^{(\mathrm{l})}[\rho(\bm{r})]$
and $v_{\mathrm{H}}^{(\mathrm{l})}[\rho(\bm{r})]$, through the canonical spinors obtained by
its diagonalization. Thus, the solutions
$\bm{c}^{(T)}$ are solved self-consistently.

The total electron density is obtained according to
\begin{equation}\label{eq:tot_el_den}
\rho(\bm{r}) = \sum_{T} \sum_{\mu, \nu} D_{\mu \nu}^{(TT)} \rho_{\mu \nu}^{(TT)}(\bm{r})
\end{equation}
where $D_{\mu \nu}^{(TT)}$, with $T=L,S$, are the diagonal blocks of the
density matrix, $\vD$, defined as
\begin{align}\label{eq:D}
\bm{D}                &= \begin{bmatrix}
                      \bm{D}^{(LL)}      & \bm{D}^{(LS)}  \\
                                   \bm{D}^{(SL)}      &  \bm{D}^{(SS)}
                     \end{bmatrix}
\end{align}
The four blocks  are defined as follows
\begin{equation}
D^{(TT')}_{\mu \nu} = \sum_a c^{(T)*}_{\mu i} c^{(T')}_{\nu i}
\end{equation}
with $TT'=LL, LS ,SL, SS$ and the the sum runs over the occupied positive-energy states.

The total energy of the electronic system is given as a functional of $\rho(\bm{r})$ and $\bm{D}$
\begin{equation}\label{tote}
        E_{tot}[\bm{D}] = Tr(\bm{D}\bm{H}_{DKS}) - 1/2 Tr(\bm{D}\bm{J})
+E_{xc}^{(l)}[\rho({\bf r})] - Tr(\bm{D}\bm{K})
\end{equation}
We have previously implemented the NOCV density partitioning within BERTHA using
the new  Python API PyBERTHA, which has provided a convenient framework to lower the
barrier to developments in relativistic quantum chemistry.  Here, we
have used a similar strategy, so that all the quantities required for the
implementation of the EDA-NOCV scheme are made available on the Python side
as NumPy arrays that can be processed efficiently.
In particular, we developed a Python function based on the {\it berthamod} module of
PyBERTHA that, for a given density matrix
($\bm{D}$), furnishes  the corresponding Dirac-Kohn-Sham matrix ($\bm{H}_{
DKS }[\bm{D}$]) and energy $E_{tot}[\bm{D}]$.
These are the sole procedures required for the implementation of the EDA-NOCV scheme.

The procedure is summarized as following.
In step 1 of the EDA scheme, we need the DKS energies and densities of the
isolated fragments A and B, which are obtained by two separate SCF
calculations of fragments A and B.  These fragments are assigned to the
respective eigenvector matrices
\begin{eqnarray} \bm{C}_{A}
= \begin{pmatrix} \bm{c}^{(L)}_A \\ \bm{c}^{(S)}_A \end{pmatrix}
& & \bm{C}_{B} = \begin{pmatrix} \bm{c}^{(L)}_B \\ \bm{c}^{(S)}_B
\end{pmatrix}
\end{eqnarray}
where the dimensions of $\bm{C}_{A}$ and $\bm{C}_{B}$ are $N^{occ}_A \cdot N_{dimA}$
and $N^{occ}_B \cdot N_{dimB}$, respectively, with $N^{occ}_{A(B)}$ being the number
of occupied positive energy states (i.e., electrons) of the system $A(B)$
and $N_{dimA(B)}$ the dimension of the G-spinor basis set for the system $A(B)$.
The corresponding density matrices are easily built as $\bm{D}_A = \bm{C}_{A} \bm{C}_{A}^\dagger$ and
$\bm{D}_B = \bm{C}_{B} \bm{C}_{B}^\dagger$ and are available as NumPy arrays.

In step 2, we need to construct the density matrix associated with
the two non-interacting fragments.  We construct the matrix $\bm{C}_+$
whose columns contain the coefficients representing in the basis set
the spinors belonging to the isolated fragments ($\vc_A^{(T)}$ and
$\vc_B^{(T)}$ for fragments A and B, respectively). The column index runs
over the occupied spinors of the isolated fragments and ranges from 1 to
$N^{occ}_A + N^{occ}_B$.  Each column is internally ordered according
to the definition of the G-spinor basis set as implemented in BERTHA,
so that the coefficients are divided into two groups: those belonging to
the large component ($T = L$) and those belonging to the small component
($T = S$).
In other words, $\vC_+$ is constructed by assembling together the sub-matrices
$\vc^{(L)}_A$, $\vc^{(L)}_B$, $\vc^{(S)}_A$, and $\vc^{(S)}_B$ with an
equal number of zero matrices as follows: \begin{equation}\label{eq:c+}
\bm{C}_{+} = \begin{pmatrix} \quad & \bm{c}^{(L)}_A & \bm{0}\\ \quad &
\quad & \quad \\ \quad & \bm{0} & \bm{c}^{(L)}_B \\ \quad & \quad &
\quad \\ \quad & \bm{c}^{(S)}_A & \bm{0} \\ \quad & \quad & \quad \\
\quad & \bm{0} & \bm{c}^{(S)}_B \end{pmatrix}
\end{equation}
The associated density matrix is obtained as matrices product $D_+ = \bm{C}_{+} \bm{C}_{+}^\dagger$. The latter
defines also the one-electron density associated at this state that is
$\rho_A+\rho_B= \sum_{T} \sum_{\mu, \nu} D_{\mu \nu}^{+ (TT)} \rho_{\mu \nu}^{(TT)}(\bm{r}) $.
The corresponding total  energy is evaluated using the functional in Eq.~\ref{tote}, where
$\rho_A+\rho_B$ and $\bm{D}_+$ is used instead of $\rho$ and $\bm{D}$, respectively.

In step 3, the new orthonormalised orbitals $\vc^{0(T)}$ are defined via the L\"owdin orthonormalisation procedure
given by
\begin{equation} \vc^{0(T)} = \vc^{(T)}_+\vO_+^{-1/2}
\end{equation}
where $\vO_+$ is the orbital spinor overlap matrix
\begin{equation}
\vO_+ = \vC_+^{\dagger}\vS\vC_+.
\end{equation}
The associated density matrix is given by $\bm{D}^0 = \bm{C}^{0} (\bm{C}^{0})^{\dagger}$
and the orthonormal electron density is
\begin{equation}
\rho^0(\bm{r}) = \sum_{T} \sum_{\mu, \nu} D_{\mu \nu}^{0 (TT)} \rho_{\mu \nu}^{(TT)} (\bm{r})
\end{equation}
Now, the total energy is evaluated using again Eq.\ref{tote}, where
$\rho^0$ and $\bm{D}^0$ are used instead of $\rho$ and $\bm{D}$, respectively.
The last step is the total orbital interaction ($\Delta E_{orb}$) which
is given by the difference between the
fully converged DKS solution and the calculation using the unrelaxed
orthonormalised electron density $\Delta E_{orb} = E[\vD]-E[\vD^0]$.
Using the NOCV spinors definition, $\Delta E_{orb}$ can be easily partitioned
in analogy with the non-relativistic case described in the previous section
and the NOCVs can be determined with the same strategy already implemented
in PyBERTHA \cite{Belpassi2020}.
The four-component generalisation of the chemical valence operator
is simple and given by
\begin{equation} \hat{V} = \sum_i \left( |
{\bm \Psi}_i^\AB \rangle \langle {\bm \Psi}_i^\AB | -| {\bm \Psi}_i^0
\rangle \langle {\bm \Psi}_i^0 | \right) \; 
\end{equation}
where  $|{\bm \Psi}_i^\AB \rangle$ and $| {\bm \Psi}_i^0
\rangle$ are the spinors being occupied in the
DKS determinant of abduct and promolecule, respectively.
The eigenstates of this operator are the relativistic NOCVs and  are clearly four-component vectors
\begin{equation}
\hat{V} {\bm \Phi_k} = \nu_k {\bm \Phi_k}
\label{eq:eigen}
\end{equation}
which, just as in the non-relativistic context, allow to put $\Delta \rho$ into a diagonal form.
In an algebraic approximation the NOCVs are the solutions of the generalised eigenvalue problem,
\begin{equation}
\bm{V} \bm{z}_k =v_k \vS \bm{z}_k
\end{equation}
where the matrix representation of the chemical-valence operator in the context of DKS is defined as
\begin{equation}
\bm{V} = \vS\Delta\vD\vS
\end{equation}
where $\Delta\vD$ is $\vD^{ AB } - \vD^0$ and $\vS$ is the  G-spinor overlap matrix given above in Eq.\ref{ss}.
Explicitly,
\begin{align}
\bm{V} = & \begin{bmatrix} S^{(LL)} \Delta D^{(LL)}S^{(LL)} & S^{(LL)} \Delta D^{(LS)} S^{(ss)} \\ S^{(SS)} \Delta D^{(SL)} S^{(LL)} & S^{(SS)} \Delta D^{(SS)}S^{(SS)} \end{bmatrix}
\end{align}
where $\Delta D^{(TT')}= D^{(TT')} -  D^{0(TT')}$.

 $\Delta\vD$ can be partitioned into contributions associated with the k-th NOCV-pairs where $\Delta\vD = \sum_k \Delta\vD^{(k)}$ with
$\Delta\vD^{(k)}=v_k (\vz_k\vz_k^\dagger - \vz_{-k}\vz_{-k}^\dagger)$.
Explicitly, $\bm{\Delta D}^{(k)}$ is defined as
\begin{align}\label{eq:H_DKS }
        \bm{\Delta D}^{(k)} &= v_k \begin{bmatrix} \bm{\Delta D}^{(k)(LL)} & \bm{\Delta D}^{(k)(LS)} \\ \bm{\Delta D}^{(k)(SL)} & \bm{\Delta D}^{(k)(SS)} \end{bmatrix}
\end{align}
where $\Delta D^{(k)(TT^{'})}= v_k(z^T_k (z^{T'}_k)^{\dagger} - z^T_{-k} (z^{T'}_{-k})^{\dagger})$.

Thus, analogously to the non-relativistic or two-components framework described in the previous section, the electron density change relate with the
orbital relaxation can be  portioned in sum of contributions associated at the k-th NOCV pair defined as:
\begin{eqnarray}\label{eq:drhok}
\Delta \rho_k(\bm{r}) & = & \sum_T\sum_{\mu \nu} \Delta D_{\mu \nu}^{{(k)} (TT)} \rho_{\mu \nu}^{(TT)}(\bm{r})
\end{eqnarray}

The total orbital energy can be also partitioned as following
\begin{eqnarray}
\Delta E_{orb}  & =    & E[\rho]-E[\rho^0] \nonumber \\
                & =    & \sum_{\mu,\nu} \Delta D_{\mu \nu} \vH^{DKS}_{\mu\nu}[\rho_{orb}^T] +O(\Delta D^3) \nonumber \\
                & =    &  \sum_k \sum_{\mu,\nu} \Delta D^k_{\mu \nu} \vH^{DKS}_{\mu\nu}[\rho_{orb}^T] +O(\Delta D^3) \nonumber \\
                & =    &  \sum_k E^{k} +O(\Delta D^3)
\end{eqnarray}
with $E^{k} \sim \sum_{\mu,\nu} \Delta D^k_{\mu \nu} \vH^{DKS}_{\mu\nu}[\rho_{orb}^T]$, and $\rho^T = 1/2(\rho +\rho^0)$. It is worth noting that while only the diagonal blocks ($\Delta D^{LL}$ and $\Delta D^{SS}$) of $\Delta\vD$ contribute in Eq.\ref{eq:drhok} to the definition of $\Delta \rho_k(\bm{r})$,
all blocks ($\Delta D^{LL}, \Delta D^{LS}, \Delta D^{SL}, \Delta D^{SS}$) are now required to define the energetic contribution, $E^k$.

We mention that in our implementation of the EDA-NOCV scheme we have used some basic features
of our new BERTHA Python API, PyBERTHA (and its associated
module \textbf{pyberthamod}, which is licensed under GPLv3, see
Ref.~\citenum{pyberthagitrefrel}, for additional and technical details
see also Refs.~\citenum{DeSantis2020a,parcopaper,Belpassi2020}).
The developed Python programme \textbf{py\_eda\_nocv.py} in which we have implemented
the EDA-NOCV scheme we have just described and the related
python modules are freely available under the GPLv3 licence at Ref.~\citenum{pyberthagitrefrel}.  A data-set collection of computational
results including numerical data, parameters and job input instructions
used to obtain the results of Section \ref{sec:results} are available
and can be freely accessed via the Zenodo repository, see Ref.\cite{belpassi_leonardo_2023_8083284}.

\section{Results and discussion}\label{sec:results}

As we have mentioned above, we have tested the validity of our new implementation on two simple cases: i) hydrogen bond in water dimer and ii) coordination bond
in the Ag$^+$-alkyne system, where relativity effects (and in particular the
spin-orbit coupling) are assumed to play
a negligible or marginal role.  After giving all relevant computational
details in Section \ref{compdet}, we will present our results in comparison
with those obtained using the ADF Modelling Suite~\cite{ADF2017authors}.
We then apply the EDA-NOCV scheme on a series of complexes
of group 6 elements, namely (CO)$_5$TM-C$_2$H$_4$, with TM =Cr, Mo, W, Sg,
where the relativistic effects play an increasingly important role
going down the group.

We will demonstrate the effective ability of the EDA-NOCV method, extended to
the full four-component DKS framework, to provide a detailed picture of the bonding and a quantitative  measure (also in
combination with the CD analysis) of the donation and back-donation components of coordination bonds, even when relativistic effects and spin-orbit coupling are crucial for describing the chemical bonding when heavy and superheavy atoms are involved.
\subsection{Computational details}\label{compdet}
All calculations were carried out using the Dirac-Kohn-Sham method
as implemented in the code BERTHA (with its new Python API,
PyBERTHA)\cite{belpassi11_12368,parcopaper,Belpassi2020}.
Density fitting techniques, which require  auxiliary basis sets, are employed in order to speed-up the calculations~\cite{belpassi06_124104,Belpassi2020}.
In all cases, the EDA-NOCV analysis was
carried out using reference fragments with even numbers of electrons. A finite charge distribution model is used for the nuclei.\cite{quiney02_5550}

In the case of the water dimer, we have used a large component
uncontracting the Dunning's double- and triple-$\zeta$ quality
basis sets\cite{Dunning1989} (denoted aug-cc-pvdz and aug-cc-pvtz and
available on the Basis Set Exchange Site\cite{basissetsite}).  The large
component of the basis set for Ag ($28s20p13d7f3g$), C and H for the
calculation of Ag$^+$-ethyne was generated by uncontraction of Dyall's
basis sets of double and triple $\zeta$-quality (aae2z and aae3z)
\cite{dyall04_403,dyall10_97,dyall02_335,dyall06_441,dyall09_12638}
which have been expanded with the corresponding polarisation and
correlation functions.  Also included in these basis sets are diffuse
functions for the s, p and d elements optimised for the anion or
extrapolated from neighbouring elements where the anion is unbound or
weakly bound.\cite{dyall:zenodo}

For the complexes of group 6 elements (CO)$_5$TM-C$_2$H$_4$, with TM
=Cr, Mo, W, Sg, we have used large component basis functions of Dyall's triple
$\zeta$-quality basis sets (aae3z) for all atoms as available in the
Zenodo repository \cite{dyall:zenodo}. This basis set corresponds to a
large component of 33s30p21d14f7g2h functions for the Sg atom.  In all
cases, the corresponding small component basis was generated using the
restricted kinetic balance relation.\cite{grant00_022508}

For the elements H, C, O and Ag, accurate auxiliary basis sets were
generated using a simple procedure derived from available DGauss Coulomb
fitting~\cite{basissetsite} basis sets (referred to as dgauss-a1-dftjfit
and available at Ref.\cite{basissetsite}).  We recall here that we use Hermite Gaussian Type Functions (HGTFs) as fitting
functions grouped into sets with the same exponents (an analogous scheme
is used in the non-relativistic DFT code DeMon~\cite{demoncf}). Due to
the variational nature of the density fitting procedure, we achieved a
fitting basis of high accuracy by simply shifting the angular momentum
of all definitions of the dgauss-Coulomb fitting upwards by two units.
For illustration, we give the dimensions of the Ag atom auxiliary basis
set, which consists of (10s,8p,8d,5f,5g).  The BP86 functional, defined
as the Becke 1988 (B88) exchange \cite{becke88_3098} plus Perdew 1986
for the correlation (PB86)\cite{PB86}, have been used. All calculations
were performed within the framework of the so-called "density-only"
DKS,\cite{jacob12_3661}. An energy convergence criterion of $10^{-8}$
Hartree was applied to the total energy.  Finally, the geometries of all
analysed systems are given in SI. These were obtained by
full geometry optimisations at the
BP86/TZ2P level using the ADF code with the ZORA scalar Hamiltonian~\cite{ADF2017authors}.

\subsection{Validation on hydrogen and coordination bonding: water dimer and Ag$^+$-C$_2$H$_2$}

The focus of this section is to prove the correctness of our new implementation.  For this
very reason, we start with a simple molecular system, the water dimer, for
which relativistic effects are negligible, and compare our numerical
results with those obtained using the EDA-NOCV implementation available
in the ADF package. As interacting fragments for the analysis we choose the two water molecules. The results have been carried out with different basis sets of increasing size and are reported in Table~\ref{tab:waterdimer}.
In the ADF code, we have used the TZ2P and QZ4P Slater- type set from the
ADF library\cite{TeVelde2001}.  Both the total interaction energy and
all other interaction energy terms ($\Delta E_{Pauli}$, $\Delta E_{Elect}$
and $\Delta E_{orb}$) show a very good agreement between the two implementations.
In particular,
when we consider the values obtained with the most accurate basis sets
(namely, QZ4P for ADF and AVTZ in
BERTHA) we found  almost identical numerical values, differences between the two
implementations always less than 0.2 kcal/mol. A very satisfactory agreement
is found also for all separated energy components  ($\Delta E^k$)  which split the orbital interaction
and are associated with NOCV pairs and for also the
 numerical values of the related eigenvalues (reported in
 parentheses also in the table).
\footnote{Note that the number of NOCV pairs is equal to the number (N)
of occupied spin orbitals, which is twenty in the case of the water dimer.
For closed-shell systems, the $\alpha$ and $\beta$ spin orbitals provide exactly the same contribution, which is implicitly accounted for in the ADF implementation by doubling the eigenvalue and by denoting the NOCV pairs from 1 to the number of spatial orbitals (N/2), avoiding any reference to spin.
In BERTHA these degenerate contributions appear
 naturally as Kramers pairs, which for simplicity are summed up for direct comparison with the ADF results.}

\begin{table}
\tbl{\label{tab:waterdimer} EDA-NOCV results of the water dimer obtained using ADF and PyBERTHA.
Energies are in kcal/mol. In PyBERTHA we use uncontracted G-spinor basis set obtained
from the aug-cc-pVDZ and aug-cc-pVTZ basis set(referred as AVDZ and AVTZ, see text for further details).
In parenthesis the eigenvalues associated with a particular NOCV pairs are reported.}
	{\begin{tabular}{lcc|cc}
\hline

                     & \multicolumn{2} {c} {ADF}& \multicolumn{2} {c} {BERTHA}  \\
\hline
Basis set            & TZP          & QZ4P                & AVDZ                         & AVTZ   \\
\hline
$\Delta E_{int}$    & -4.66        &  -4.35              &  -4.46                       & -4.33   \\
$\Delta \tilde E_{Pauli}$   & -        &   -                 &  14.10                       & 13.93   \\
$\Delta E^0_{XC}$    &   -          &   -                 &  -5.07                       & -5.08   \\
$\Delta E_{Pauli}$ & 8.58         &   8.98              &   9.03                       &  8.85   \\
$\Delta E_{elstat}$ & -8.63        &  -8.87              &  -9.28                       & -8.99   \\
$\Delta E_{orb}$   & -4.66        &   -4.46             &  -4.20                       & -4.20   \\
$\Delta E^{1}$& -3.85 (0.1369) & -3.82 (0.1368)   &  -3.54(0.1321)               & -3.61(0.1328)   \\
$\Delta E^{2}$& -0.37 (0.0393) & -0.22 (0.0310)   &  -0.23(0.0307)               & -0.19(0.0300)   \\
$\Delta E^{3}$& -0.11 (0.0217) & -0.15 (0.0270)   &  -0.15(0.0264)               & -0.14(0.0260)   \\
$\Delta E^{4}$& -0.06 (0.0191) & -0.11 (0.0234)   &  -0.10(0.0251)               & -0.10(0.0252)   \\
$\Delta E^{5}$& -0.07 (0.0166) & -0.07 (0.0177)   &  -0.08(0.0178)               & -0.07(0.0176)   \\
$\Delta E^{6}$& -0.08 (0.0143) & -0.05 (0.0153)   &  -0.06(0.0161)               & -0.06(0.0155)   \\
\hline
	\end{tabular}
	}
\end{table}
In Table \ref{tab:ag} we present a numerical comparison for the
Ag$^+$-ethyne complex as an example of coordination bond
(Ag$^+$ and alkyne are the interacting fragments).
In particular, we report the EDA-NOCV analysis using both ADF (at ZORA scalar
and non-relativistic Hamiltonian levels and using the QZ4P Slater
basis set) and our new implementation in BERTHA using Dyall's triple
$\zeta$ basis set.  We can expect that the effect of spin-orbit coupling
is negligible here ($\Delta E_{int}$ evaluated using
ZORA-Spin Orbit Hamiltonian in ADF is of -39.33 kcal/mol) and that the system can be described well by including the relativistic effects at scalar level . All data
obtained using the Zora
Scalar Hamiltonian in ADF agree with those obtained
with BERTHA at 4-component DKS level.  We have also carried out the analysis by increasing the speed of light in BERTHA by 2 orders of magnitude
(i.e. c = 13703.6~a.u.) to approach the non-relativistic limit and we
found a satisfactory agreement with the non-relativistic results of ADF (labelled as n.r. in the Table).
Summarising, all the results
obtained with the EDA-NOCV implementation of the ADF and those obtained using BERTHA are in very good agreement. Thus, the small differences are within the numerical variation due to the inequalities in the basis sets employed. This
makes us confident that our implementation is both numerically stable and correct.

Before concluding this Section, it is interesting to point out that
the term $\Delta E^0_{XC}$,
usually combined with $\Delta \tilde E_{Pauli}$
 to give the total Pauli term ($\Delta E_{Pauli}$)
is far from being negligible ( $\Delta E^0_{XC}$ is about 30-40\% of $\Delta \tilde E_{Pauli}$ for both
hydrogen and coordination bonds) and it is actually a significant fraction of $\Delta E_{Pauli}$. This interesting finding deserves further investigation, particularly because it is often overlooked in the literature.

\begin{table} 
\tbl{\label{tab:ag} EDA-NOCV results of the Ag$^+$-alkyne system obtained using ADF at ZORA scalar and non relativistic (n.r.) level and PyBERTHA using two different speed of light $c$.
Energies are in kcal/mol. In PyBERTHA we use uncontracted G-spinor basis set obtained
from the Dyall's triple-$\zeta$ basis set while in ADF the Slater QZ4P has been used.
In parenthesis the eigenvalues associated with the NOCV deformation densities are reported.}
	{\begin{tabular}{lccccc}
\hline
\hline
                         &  \multicolumn{2} {c} {ADF}       & & \multicolumn{2} {c} {BERTHA}    \\
                         &                  &               & &                  &   \\
                         &  ZORA Scalar     & n.r.          & &   c=137.036      &  c=13703.6 \\
        \cline{2-3}\cline{5-6}
                         &                  &               & &                  &   \\
$\Delta E_{int}$       &  -39.23          & -31.2         & &    -39.31        & -31.20           \\
$\Delta \tilde E_{Pauli}$ &   -              &  -            & &     83.94        &  84.50           \\
$\Delta E^0_{XC}$        &   -              &  -            & &    -29.13        & -28.43           \\
$\Delta E_{Pauli}$     &  54.57           & 55.83         & &     54.82        &  56.07           \\
$\Delta E_{elstat}$     & -53.54           &-52.88         & &    -53.56        & -52.90           \\
$\Delta E_{orb}$       &  -40.26          &-34.21         & &    -40.57        & -34.37           \\
$\Delta E^{1}$         &  -24.27 (0.5011) &-19.25 (0.4246)& &    -24.37(0.5020)& -19.27(0.4254)   \\
$\Delta E^{2}$         &  -7.09  (0.2479) &-6.59  (0.2284)& &     -7.22(0.2513)&  -6.67(0.2306)   \\
$\Delta E^{3}$         &  -3.53  (0.1326) &-3.47  (0.1308)& &     -3.56(0.1337)&  -3.49(0.1315)   \\
$\Delta E^{4}$         &  -2.58  (0.1008) &-2.37  (0.0937)& &     -2.58(0.1015)&  -2.36(0.0943)   \\
\hline
\hline
\end{tabular}
}
\end{table}

\subsection{Complexes of group 6 elements (CO)$_5$TM-C$_2$H$_4$, with TM=Cr, Mo, W and Sg.} \label{goldand}

The coordination bonding of ethylene to transition metals
(TM) has been extensively studied, and the bonding is
generally described using the DCD bonding model mentioned earlier
\cite{Dewar:51,Chatt:53,frenking2001understanding,C5SC02971F}. The interaction between the metal and ethylene results from the $\sigma$-donation of electron charge from the
occupied orbitals of ethylene to the empty metal orbitals, and by a $\pi$-back-donation from the occupied orbitals of the metal  to the empty orbitals of ethylene (see Ref. \cite{C5SC02971F} and the references therein for
a detailed discussion of DCD model).

The bonding situation between ethylene and group 6 metal carbonily complexes (CO)$_5$TM-C$_2$H$_4$ (TM =Cr, Mo and W) was previously
investigated by Frenking et al.~\cite{Frenking:2004} using EDA
including scalar relativistic effects within ZORA
approximation. Here we extend the series to the 7th period and analyse
alkene-metal bonding in the entire series of group 6 (CO)$_5$TM-C$_2$H$_4$
complexes including the superheavy element Sg (Z=106). For completeness we also report  the results for the whole series obtained using the ZORA scalar Hamiltonian in SI (Table S)

\begin{table}
        \tbl{\label{tab:TM}EDA-NOCV results for the  (CO)$_5$TM-C$_2$H$_4$ complexes with TM=Cr, Mo, W and Sg. Energy values in kcal/mol.
        Data obtained at DKS level using BERTHA in combination with Dyall's aae3Z basis set, see text for details.
        Charge transfer (CT) in electrons are extracted from the charge-displacement analysis, values of CD function at the isodensity boundary (see text for details).
        Negative (positive) value corresponds to an electron charge transfer in the direction going from the metal (ethylene) fragment to the
	ethylene (metal) fragment.}
	{
\begin{tabular}{lcccc}
                                  &  \multicolumn{4} {c} {(CO)$_5$TM-C$_2$H$_4$}            \\
\hline
                                  &   Cr(CO)$_5$-C$_2$H$_4$ &  Mo(CO)$_5$-C$_2$H$_4$ &  W(CO)$_5$-C$_2$H$_4$  &  Sg(CO)$_5$-C$_2$H$_4$      \\
\hline
                                  &                   &                    &                     &                             \\
        $\Delta E_{int}$          &   -29.43           &  -28.34           & -34.72            &          -33.30         \\
        $\Delta \tilde E_{Pauli}$          &   128.72           &  115.10           & 141.58            &          147.56         \\
        $\Delta E^0_{XC}$           &   -46.65           &  -41.25           & -43.56            &          -48.42         \\
$\Delta E_{Pauli}$              &    82.07           &   73.85           &  98.02            &           99.13         \\
$\Delta E_{elstat}$              &   -57.98           &  -55.30           & -75.04            &          -74.11         \\
$\Delta E_{orb}$                &   -53.52           &  -46.89           & -57.70            &          -58.33         \\
        $\Delta E^{1}$          &   -21.93(0.5746)   &  -20.02(0.5604)   & -25.50(0.6201)    &          -26.47(0.6349)   \\
        $\Delta E^{2}$          &   -25.78(0.5495)   &  -21.54(0.4723)   & -25.67(0.4996)    &          -25.30(0.4937)   \\
        $\Delta E^{3}$          &    -1.76(0.1059)   &   -1.39(0.0987)   &  -1.60(0.1135)    &           -1.76(0.1107)   \\
        $\Delta E^{4}$          &    -1.58(0.0919)   &   -1.42(0.0870)   &  -1.79(0.0968)    &           -1.66(0.0917)   \\
                                  &                    &                   &     &                             \\
        CT$_{tot}$                &  -0.030            &  -0.0427          &  -0.053   &   -0.038        \\
        CT$_{1}$                  &  -0.244            &  -0.234           &  -0.260   &   -0.231        \\
        CT$_{2}$                  &   0.226            &   0.200           &  0.211   &    0.193        \\
        CT$_{3}$                  &   -0.007           &  -0.010           &  -0.009   &   -0.008        \\
        CT$_{4}$                  &   0.004            &   0.008           &  0.009   &    0.007        \\
\end{tabular}
	}
\end{table}

\begin{figure}[htb]
  \includegraphics[width=1.\textwidth]{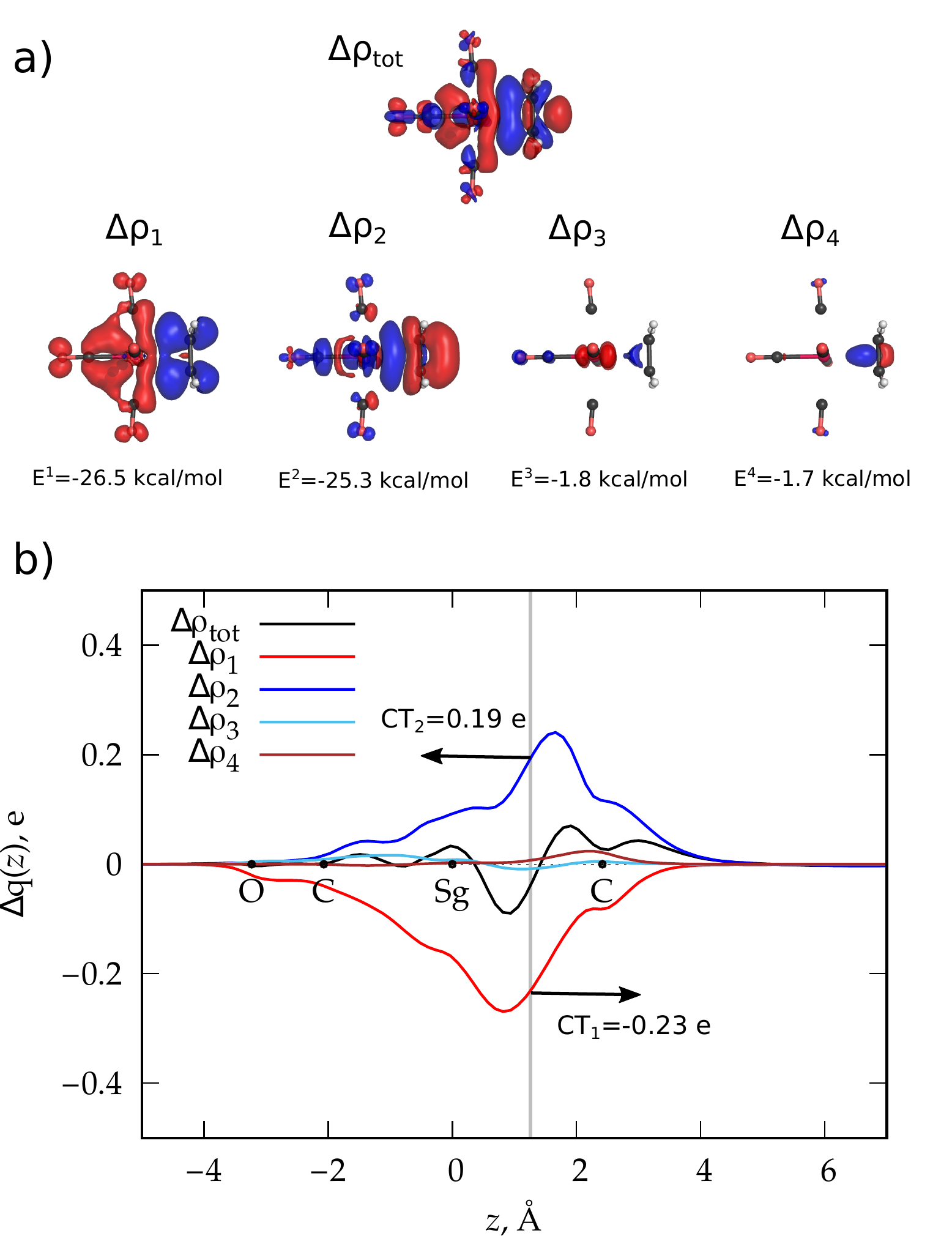}
  \caption{NOCV-CD analysis for the Sg-ethylene bond in the
  (CO)$_5$Sg-C$_2$H$_4$ complex. a) Contribution to the total deformation density,
$\Delta \rho$, of the four most significant NOCV-pairs ($\Delta \rho_1$,
$\Delta \rho_2$,
$\Delta \rho_3$ and $\Delta \rho_4$,).Isodensity
surfaces ($\pm 0.0014$ e a.u.$^{-3}$) for $\Delta \rho$.
Red surfaces identify charge depletion areas, blue surfaces identify charge accumulation areas. b) CD curves. The vertical line marks the boundary between the (CO)$_5$Sg and
the alkene fragment (see text for details).
 The dots on the axis mark the z coordinate of the atoms.
 }
\label{cd:sg}
\end{figure}

The results of our analysis are given in Table~\ref{tab:TM} for all
complexes. The Table also lists the CT values determined with the NOCV/CD approach.
As an illustrative example we show in Figure \ref{cd:sg}, here for
the Sg(CO)$_5$-C$_2$H$_4$ system, the total electron density rearrangement
($\Delta \rho$) that occurs in the bond formation between the metal
fragment (CO)$_5$Sg and ethylene, and the most significant NOCVs
pair decomposition ($\Delta \rho_k$, with $k=1,2,3,4$). In the Figure are reported the
isosurface plots together with the corresponding CD functions (see Section \ref{compdet}),
which provide quantitative information about the charge shift that
actually occurs during bond formation. Similar results are reported for
the lighter homologoues in the SI (see Figures~S1, S2, S3 for Cr, Mo and W, respectively).

The numerical data show an interaction
energy that follows the trend Mo < Cr < Sg <W with the superheavy
element Sg interaction energy only slightly smaller (less
than 2 kcal/mol) than the one of W.
The overall picture that emerges from our analysis is that the
metal-ethylene bonding has a very similar
character along the group. As already observed by Frenking et al. for Mo, Cr and W, the
metal-substrate bond appears to be more electrostatic rather than covalent,
although the orbital energy is an important attractive component of
the interaction (the ratio $\Delta E_{elstat}/\Delta E_{orb}$ is always greater than one, i.e. 1.08, 1.18, 1.30 and 1.27 for Cr, Mo,
W and Sg, respectively).

A more stringent comparison
between W and Sg shows that not only the total interaction energy is very
similar (-34.7 kcal/mol and -33.3 kcal/mol for W and Sg, respectively), but also all the other energy contributions ($\Delta E_{Pauli}$, $\Delta
E_{elstat}$ and $\Delta E_{orb}$) appear to have very similar numerical values, which differ by less than 2 kcal/mol.
These systems exhibit also the largest orbital interaction (-57.7 and -58.3 kcal/mol
for W and Sg, respectively) among the series.
The NOCVs decomposition of the total orbital
interaction ($\Delta E_{orb}$) clearly shows that there are two main contributions of comparable strength ($\Delta E^1$ and $\Delta E^2$).
The reader can easily understand the interpretive power of
the EDA-NOCV approach, which can associate
these stabilising contributions with respect to the corresponding NOCV electron density deformations. In particular, here we can easily identify the
character of these two components ($\Delta E^1$ and $\Delta E^2$) thanks to the simple visual inspection
of the isodensity-surface plots of the associated deformation densities
(see $\Delta \rho_1$ and $\Delta \rho_2$ for the specific case of Sg in Figure \ref{cd:sg}). The analysis for the other systems is reported in the SI.

Despite the fact that total electron density rearrangement ($\Delta
\rho$ ) shows a very complex pattern of charge accumulation (blue
isosurface) and charge depletion (red isosurface),  the NOCV pair
deformation densities can be clearly characterised. In particular, $\Delta \rho_1$
is characterised by
depletion at the site of the metal fragment with an evident accumulation
on ethylene. Clearly, this
represents the backdonation component with a charge flux going from the metal fragment to the
unoccupied in-plane $\pi$ orbitals of ethylene.
The second deformation density ($\Delta \rho_2$)
 is inversely characterised by charge depletion at the ethylene site
 and charge accumulation at the metal fragment and it can be seen as the finger print of the donation component of the DCD model. Note that the accumulation of electron density is not confined to the
metal, but involves also the CO ligands with a
pattern of charge rearrangements that depends on the specific position of the CO ligand in the fragment. In the panel b) of Figure \ref{cd:sg},
we also show the NOCV-CD analysis for the Sg complex.
Recall that at a given point along a chosen axis (here the
axis that joins the metal center with the CC bond mid point of ethylene), $z$,
a positive value of the CD function corresponds to
a flow of electrons from the right to the left; in our case from ethylene to the metal fragment.
Conversely, a negative value of the CD function corresponds to a flow
in the opposite direction, i.e. from the metal Sg to the ethylene fragment.

A reasonable measure of the charge transfer (CT) between the ethylene and
the Sg fragment can be easily determined by setting a plausible boundary to
separate the fragments within the complex. Our standard choice is the $z$
point where equal-valued isodensity surfaces of the isolated fragments
become tangent\cite{belpassi08_1048,cappelletti12_1571,bistoni13_11599}.
The vertical black line in the figure marks this ''isodensity boundary''.
Thus, using the NOCV/CD analysis, we can also give a quantitative picture of the basic binding
modes in term of CT associated with the components of the DCD bonding model while represent a complementary information of the EDA-NOCV energy partition. The overall CD function clearly results mainly from the metal to ethylene back-donation component (labelled as $\Delta \rho_1$, red curve in the
diagrams), which is large and negative and a
second component (labelled $\Delta \rho_2$, the blue line and the blue
dashed line, respectively) which is positive and clearly quantifies  the
ligand to metal donation. Quantitatively, the CT associated at the backdonation (CT=-0.23 e) is larger, in absolute value, than that related with the donation component (CT=0.19 e). Such pattern is common among all complexes. Our EDA-NOCV analysis (and NOCV-CD) shows that the W and Sg complexes have both the largest orbital interaction with very
similar NOCVs energy components and associated CTs.
Our analysis shows that in this series of complexes
Sg possesses a bond with ethylene which is very similar to that of W and characterized by similar DCD components (donation and back-donation) that are even
larger than those of the lighter homologues. The Sg-ethylene bond is characterized by a significant back-donation component both in terms of  the large energy stabilization ($\Delta E^1=$ -26.5 kcal/mol, see Table\ref{tab:TM} ) and CT (-0.23 e, see Table\ref{tab:TM} and Figure\ref{cd:sg}). This finding is somewhat unexpected on the basis of a recent theoretical analysis\cite{Pershina:2017}
in hexacarbonyl complexes  of Mo, W and Sg, where
the slightly lower interaction energy of Sg compared to W was mainly attributed to a weaker metal to CO backdonation in Sg(CO)$_6$ than in W(CO)$_6$. It is clear that systematic studies on different complexes, including, for instance, SgO$_2$Cl$_2$ and SgO$_2$(OH)$_2$ which were also experimentally studied, are mandatory to gain a more detailed picture of the bonding properties of Sg.

\section{Conclusions}
In the present work we have extended the EDA-NOCV
 method to the relativistic four-component DKS framework. This method
 allows us to analise  a chemical bond in
 terms of well-defined energy components ($ \Delta E_{int} = \Delta E_{elstat} +
\Delta E_{Pauli} + \Delta E_{orb} $). It provides a consistent energy/charge
 splitting scheme, where the different contributions to the orbital relaxation can be
easily identified by visualising the associated deformation density and
quantified by providing the corresponding stabilization energies.
Thanks to this new implementation, the approach can be now applied
to the analysis of the chemical bond in molecular systems containing heavy and super-heavy
atoms in which the relativistic effects, including spin-orbit coupling,
need to be considered at the highest level of accuracy.

This new implementation has been carried out in the framework of the DKS theory, and it has been validated
by comparing the results with those obtained using the EDA-NOCV
implemented  in the ADF package. The benchmark study was successfully performed with
different basis sets for selected systems where relativistic effects are not
expected to play a significant role.
 Finally, we have carried out a systematic analysis of
 the metal-ethylene coordination bond in the group 6-element series
 (CO)$_5$TM-C$_2$H$_4$, with TM= Cr, Mo,
W and Sg, where relativistic effects are likely to play an increasingly
important role as one moves down the group.  In particular, we
have shown that the EDA-NOCV method, in combination with charge
displacement analysis, used in the framework of four-component relativistic
calculations, is able to identify the donation and back-donation charge
fluxes of the Dewar-Chatt-Duncanson bonding model, which is ubiquitous
in coordination chemistry. Our analysis shows that the Sg complex has both very similar EDA-NOCV energetic partition and DCD components
(donation and back-donation) to those of the W complex and even larger than those of the lighter homologues.

We believe that the methodology presented in
this work, together with other advances in the
field,~\cite{Saue:2001,Reiher:2007,Sablon2010,Klobukowski:2011,Galland:2014}
may be useful to rationalise the effect of relativity, including spin-orbit coupling, on
bonding, reactivity and experimental observations in molecules containing
heavy and superheavy elements.

\section*{Acknowledgement(s)}
This work received financial support from ICSC-Centro Nazionale di Ricerca in High
Performance Computing, founded by European Union-Next-Generation-UE-PNRR, Missione 4 Componente 2 Investimento 1.4. CUP:  B93C22000620006.
Research at SCITEC-CNR has been funded by the European Union - NextGenerationEU under the Italian Ministry of
University and Research (MUR) National Innovation Ecosystem grant ECS00000041 - VITALITY. 
L. B. and P. B. acknowledge Universit\`{a} degli Studi di Perugia and MUR, CNR for support within the project Vitality.
L. B. and L. S. thank all the attendances at the REHE conference held in September 2022 in Assisi (Italy) for stimulating discussions
which inspired this work.

\begin{appendix}
\section{\label{appendix:a} Demonstration that the energy difference $E[\rho_A+\rho_B]-E[\rho_A]-E[\rho_B]$
        is given by the sum of $\Delta E_{elstat}$ and $\Delta E_{XC}^0$.}

Here we explicitly show that the energy difference between the energy associated with
the system obtained using the unmodified electronic densities of the fragments put at their final position
        in the adduct ($E[\rho_A+\rho_B]$) and that of the isolated fragments ($E[\rho_A]$
and $E[\rho_B]$) is given by the sum of the electrostatic interaction
        ($\Delta E_{elstat}$) plus an exchange-correlation term ($\Delta E^0_{XC}$).

We start with the explicit definition of $E[\rho_A+\rho_B]$ in Eq.~\ref{appeq:dim1}
\begin{eqnarray}
        \label{appeq:dim1}
        E[\rho_A+\rho_B] &=& T_{A+B} - \sum_{a \in \{A,B\}} \int v_a(r) (\rho_A(r)+\rho_B(r)) dr + \nonumber \\
                         &+& \frac{1}{2}\int \int \frac{(\rho_A(r_1)+\rho_B(r_1))(\rho_A(r_2)+\rho_B(r_2))}{|r_1-r_2|} dr_1 dr_2 +
                    E_{XC}[\rho_A+\rho_B] + \\
                         &+& \sum_{\substack{i>j \\{i,j \in \{A,B\}}}} \frac{Z_i Z_j}{|R_i-R_j|} \nonumber
\end{eqnarray}
where $T_{A+B}$ is the kinetic energy term, $v_a(r)$ is the attractive
electron-nuclei Coulomb potential for the whole system,
the third term is the electronic Coulomb repulsion,
$E_{XC}[\rho_A+\rho_B] $ is the exchange-correlation energy contribution and finally
there is the total nuclei-nuclei Coulomb repulsion term.
Now, because the orbitals (or spinors) of the two fragments are not allowed to relax in
this step of the EDA scheme (see also Fig. 1 in the manuscript), we can easily recognise that the kinetic term
$T_{A+B}$ is given by the sum of the kinetic energy of the two separated fragments ($T_{A}$ and $T_{B}$).
If we expand Eq.~\ref{appeq:dim1}, gathering togethar those terms
that are exclusively associated to one fragment (A or B), and add and subtract
        the associated exchange-correlation energy terms ($E_{XC}[\rho_A$] and $E_{XC}[\rho_B$]),
we arrive at the following equation
\begin{eqnarray*}
 \begin{aligned}
 &\left.\begin{aligned}
        E[\rho_A+\rho_B] &=& T_{A} - \sum_{a \in \{A\}} \int v_a(r) (\rho_A(r)) dr + \frac{1}{2}\int \int \frac{\rho_A(r_1)\rho_A(r_2)}{|r_1-r_2|} dr_1 dr_2 +  \\
                     &+& \sum_{\substack{i>j \\{i,j \in \{A\}}}} \frac{Z_i Z_j}{|R_i-R_j|}+E_{XC}[\rho_A]+\\
\end{aligned}\hspace{1.5cm}\right\}\hspace{0.5cm}\text{$E[\rho_A]$}\\
&\left.\begin{aligned}
        \hspace{2cm}            &+& T_{B} - \sum_{a \in \{B\}} \int v_a(r) (\rho_B(r)) dr + \frac{1}{2}\int \int \frac{\rho_B(r_1)\rho_A(r_2)}{|r_1-r_2|} dr_1 dr_2 +  \\
\
        &+& \sum_{\substack{i>j \\{i,j \in \{B\}}}} \frac{Z_i Z_j}{|R_i-R_j|}+E_{XC}[\rho_B] - \\
\end{aligned}\hspace{1.5cm}\right\}\hspace{0.5cm}\text{$E[\rho_B$]}\\
&\left.
\begin{aligned}
        \hspace{2cm}       &-& \sum_{a \in \{A\}} \int v_a(r) (\rho_B(r)) dr -\sum_{a \in \{B\}} \int v_a(r) (\rho_A(r)) dr +\sum_{\substack{i \in \{A\} \\j \in \{B\}}}
 \frac{Z_i Z_j}{|R_j-R_i|} + \\
               &+& \int \int \frac{(\rho_A(r_1)\rho_B(r_2))}{|r_1-r_2|} dr_1 dr_2 + E_{XC}[\rho_A+\rho_B] - E_{XC}[\rho_A] - E_{XC}[\rho_B]
\end{aligned}\hspace{0.1cm}\right\}\hspace{0.4cm}
\begin{array}{c}
\Delta E_{elstat} \\
        +  \\
        \Delta E^0_{XC} \\
\end{array}
\end{aligned}
\end{eqnarray*}
where the first five terms correspond to the definition of the energy of the isolated fragment A ($E[\rho_A]$)
and the following five sum up to the energy of the isolated fragment B (E[$\rho_B$]).
Finally, we write the desire energy difference as
\begin{eqnarray}
        E[\rho_A+\rho_B]   -  E[\rho_A] - E[\rho_B] = \Delta E_{elstat} + \Delta E^0_{XC}
\end{eqnarray}

where
\begin{equation}
\Delta E_{elstat}  =  \sum_{a \in \{A\}} \int v_a(r) (\rho_B(r)) dr -\sum_{a \in \{B\}} \int v_a(r) (\rho_A(r)) dr +\sum_{\substack{i \in \{A\} \\j \in \{B\}}} \frac{Z_i Z_j}{|R_j-R_i|} +\int \int \frac{(\rho_A(r_1)\rho_B(r_2))}{|r_1-r_2|} dr_1 dr_2
\end{equation}
defines the electrostatic term of EDA
and the term $\Delta E^0_{XC}$  identifies an exchange-correlation contribution
\begin{eqnarray*}
        \Delta E^0_{XC} & = & E_{XC}[\rho_A+\rho_B] - E_{XC}[\rho_A] - E_{XC}[\rho_B]
\end{eqnarray*}
which, as pointed out the the main text, is typically included
in the Pauli term.
\section{Formalism of Transition State (TS) method}\label{appendix:b}

The Transition State (TS) formalism was introduced by Slater in the early seventies
with the aim to estimate the ionization energy and electronic transitions.~\cite{SLATER19721}
Its name may generate a certain confusion for a chemist beacuse it recalls
the idea of transtion state theory, however the method is actually a simple scheme to evaluate
the energy difference between two states
expanding in series their energies around a fictisiuos one
(called transition state) settled in between.
The method has been extended by Tom Ziegler and A. Rauk to evaluate
the bonding energy within the contex of the Hartree-Fock-Slater method\cite{zieglerETS:1977}.

In the following we derive the basis equations, mainly following the original work
of Ziegler\cite{zieglerETS:1977}, using only a slighthly different notation
(based on one electron density matrix) to be consinstent with the notation used in this work.
As mentioned above, the basic idea is to expand in a Taylor series the energy of both the final state ($E[\rho_f]$)
and the initial state ($E[\rho_i]$ respect a common origin, $E[\rho_T]$,
which is the energy associated with a system with the average density ($\rho_T = 1/2(\rho_f +\rho_i$).
Analogous to the Taylor expansion of multi-variables function $f(x_1,...,x_N)$ around the point $(x^T_1,...,x^T_N)$:
\begin{equation}
f(x_1,...x_N)= f(x^T_1,...,x^T_N) +  \sum_{i=1}^N \left( \frac{\partial f(x_1,...,x_N) }{\partial x_i} \right)_{(x^T_1,...,x^T_N)} (x_1-x^T_1,...,x_N-x^T_N) + ....
\end{equation}
here we expand the Kohn-Sham (or Dirar-Kohn-Sham) energy
\begin{equation}
E_{KS}[\rho] = Tr({\bm Dh}) + \frac{1}{2} Tr({\bm D J[D]}) +E_{xc}[\rho]
\end{equation}
with respect to the one electron density matrix elements $D_{kl}$ where
$\rho(r)=\sum_{k,l} D_{kl} \chi^*_k(r)\chi_l(r)$.
For the pourpose of the derivation it is usefull
to recall that the partial derivative of the Kohn-Sham energy respect the density
matrix elements ($D_{ij}$) is given by
\begin{eqnarray}
\frac{\partial E[\rho] }{\partial D_{ij}} & = & h_{ij} + J_{ij} + \int \underbrace{\frac{\delta E_{xc}[\rho]}{\delta \rho(r)}}_{v^{xc}[\rho](r)} \frac{\partial \rho(r)}{\partial D_{ij}}dr \\
                                          & = & h_{ij} + J[\rho]_{ij} + V^{XC}[\rho]_{ij} \\
                                          & = & F[\rho]_{ij}
\end{eqnarray}
where $F[\rho]_{ij}$ is the elements of the Kohn-Sham matrix associated with the
density, $\rho(r)$.
Expanding the energy of both the final state ($E_{KS}[\rho^f]$) and the
initial state ($E_{KS}[\rho^i]$)
with respect a common stransition state energy ($E_{KS}[\rho^T]$) which corresponds to the energy of a fictisiuos
electronic state associated with the everaged density ($\rho^T=\frac{1}{2}(\rho^f-\rho^i)$
we have that energies of the two states can be written in terms of the elements of the
matrix $\Delta D$ as follows
        \begin{eqnarray*}
E_{KS}[\rho^f]  & = & E_{KS}[\rho^T] + \sum_{kl} \left(\frac{\partial E[\rho]}{\partial D_{kl}}\right)_{\rho^T} (D^f_{kl} -D^T_{kl}) + \\
                & + &  \frac{1}{2}\sum_{klij} \left(\frac{\partial^2 E[\rho]}{\partial D_{kl}\partial D_{ij}}\right)_{\rho^T} (D^f_{kl} -D^T_{kl}) (D^f_{ij} -D^T_{ij}) + O(\Delta D^3)  \\
                & = & E_{KS}[\rho^T] + \sum_{kl} F[\rho^T]_{kl} (\frac{1}{2}\Delta D_{kl}) + \frac{1}{2}\left(\frac{\partial^2 E[\rho]}{\partial D_{kl}\partial D_{ij}}\right)_{\rho^T} (\frac{1}{2}\Delta D_{kl})(\frac{1}{2}\Delta D_{ij}) + O(\Delta D^3) \\
E_{KS}[\rho^i]  & = & E_{KS}[\rho^T] + \sum_{kl} \left(\frac{\partial E[\rho]}{\partial D_{kl}}\right)_{\rho^T} (D^i_{kl} -D^T_{kl}) + \\
                & + & \frac{1}{2}\sum_{klij} \left(\frac{\partial^2 E[\rho]}{\partial D_{kl}\partial D_{ij}}\right)_{\rho^T} (D^i_{kl} -D^T_{kl}) (D^i_{ij} -D^T_{ij}) + O(\Delta D^3) \\
                & = & E_{KS}[\rho^T] + \sum_{kl} F[\rho^T]_{kl} (-\frac{1}{2}\Delta D_{kl}) + \frac{1}{2}\left(\frac{\partial^2 E[\rho]}{\partial D_{kl}\partial D_{ij}}\right)_{\rho^T} (-\frac{1}{2}\Delta D_{kl})(-\frac{1}{2}\Delta D_{ij}) + O(\Delta D^3)
        \end{eqnarray*}
In the above expressions, $D^T_{kl}$ are equal to $\frac{1}{2}(D_{kl}^f-D_{kl}^i)$.
The second order term in both expansions above cancel out in the energy difference
($E_{KS}[\rho^f]  - E_{KS}[\rho^i]$)
and ones  obtains an expression  which is correct up to the second order
and formally linear in $\Delta D$.
\begin{equation}
        \label{appeq:e2}
        E_{KS}[\rho^f]  - E_{KS}[\rho^i] = \sum_{kl} F[\rho^T]_{kl} (\Delta D_{kl}) + O(\Delta D^3) + ..  = Tr({\bm F}[\rho^T]{\bm \Delta D} + O(\Delta D^3)
\end{equation}
In their seminal work, Ziegler and Rauk went a step further, deriving even the third order correction
showing also this term can be written throught an expression which is
formally linear in $\Delta D$.
This can be easily showed considereing the third order contribution ($\Delta E(\Delta D^3)$)
\begin{eqnarray}
        \Delta E(\Delta D^3) = \frac{1}{3!} \frac{1}{4}\sum_{ij,kl,mn}  \left(\frac{\partial^3 E[\rho]}{\partial D_{kl}\partial D_{mn}\partial D_{ij}}\right)_{\rho^T} \Delta D_{kl}\Delta D_{mn}\Delta D_{ij}
\end{eqnarray}
and using $\frac{\partial E[\rho]}{\partial D_{ij}}=F[\rho]_{ij}$ to write
\begin{eqnarray}
        \label{appeq:e3}
        \Delta E(\Delta D^3) = \frac{1}{3!} \frac{1}{4}\sum_{ij} \left( \sum_{kl,mn}  \left(\frac{\partial^2 F[\rho]_{ij}}{\partial D_{kl}\partial D_{mn}}\right)_{\rho^T} \Delta D_{kl}\Delta D_{mn}\right) \Delta D_{ij}
\end{eqnarray}
The term in parenthesis in Eq.\ref{appeq:e3} can be evaluated expanding both
F$[\rho^f]_{ij}$  and F$[\rho^i]_{ij}$ in Taylor series around F$[\rho^T]_{ij}$
\begin{eqnarray*}
F[\rho^f]_{ij} & = & F[\rho^T]_{ij} + \sum_{kl} \left(\frac{\partial F[\rho]}{\partial D_{kl}}\right)_{\rho^T} (\frac{1}{2}\Delta D_{kl}) + \sum_{kl,mn} \frac{1}{2} \left(\frac{\partial^2 F[\rho]}{\partial D_{kl}\partial D_{mn}}\right)_{\rho^T} (\frac{1}{2}\Delta D_{kl})(\frac{1}{2}\Delta D_{mn}) \\
F[\rho^i]_{ij} & = & F[\rho^T]_{ij} + \sum_{kl} \left(\frac{\partial F[\rho]}{\partial D_{kl}}\right)_{\rho^T} (-\frac{1}{2}\Delta D_{kl}) + \sum_{kl,mn} \frac{1}{2} \left(\frac{\partial^2 F[\rho]}{\partial D_{kl}\partial D_{mn}}\right)_{\rho^T} (-\frac{1}{2}\Delta D_{kl})(-\frac{1}{2}\Delta D_{mn})
\end{eqnarray*}
and by summing term by term we obtain
\begin{eqnarray}
        \label{appeq:ef}
        F[\rho^f]_{ij} + F[\rho^i]_{ij} & = & 2 F[\rho^T]_{ij} +\frac{1}{4} \sum_{kl,mn} \left(\frac{\partial^2 F[\rho]}{\partial D_{kl}\partial D_{mn}}\right)_{\rho^T} \Delta D_{kl} \Delta D_{mn} \\
\end{eqnarray}
and rearranging one arrives at  Eq.~\ref{appeq:ef1}
\begin{equation}
        \label{appeq:ef1}
        \sum_{kl,mn} \left(\frac{\partial^2 F[\rho]_{ij}}{\partial D_{kl}\partial D_{mn}}\right)_{\rho^T} \Delta D_{kl} \Delta D_{mn}  =  4(F[\rho^f]_{ij} + F[\rho^i]_{ij} -2 F[\rho^T]_{ij})
\end{equation}
Using this result, the expression originally derived by  Ziegler and Rauk
and corrected
up to the fourth order in $\Delta D$ is given by
\begin{eqnarray}
        \label{appeq:corr3}
        E_{KS}[\rho^f]  - E_{KS}[\rho^i] & = & \sum_{ij} F[\rho^T]_{ij} (\Delta D_{ij}) + \frac{1}{24} \sum_{ij} \sum_{kl,mn} \left(\frac{\partial^2 F[\rho]_{ij}}{\partial D_{kl}\partial D_{mn}}\right)_{\rho^T} \Delta D_{kl} \Delta D_{mn} \Delta D_{ij}
        \nonumber \\
                                         & = & \sum_{ij} F[\rho^T]_{ij} (\Delta D_{ij}) + \frac{1}{24} \sum_{ij} (4(F[\rho^f]_{ij} + F[\rho^i]_{ij} -2 F[\rho^T]_{ij}) \Delta D_{ij} \nonumber \\
                                         & = & \sum_{ij} \left( \frac{2}{3} F[\rho^T]_{ij} + \frac{1}{6} F[\rho^f]_{ij} +\frac{1}{6} F[\rho^i]_{ij} \right) \Delta D_{ij} +  O(\Delta D^5)
\end{eqnarray}
and is formally linear in $\Delta D$.
As already pointed out in the main text, the possibility of write the energy difference ($E_{KS}[\rho^f]  - E_{KS}[\rho^i]$)
using formula (see Eq.~\ref{appeq:e2} and Eq.~\ref{appeq:corr3}),
formally linear in $\Delta D_{ij}$, is crucial in order
to use the partitioning scheme based on the NOCV method.

Before concluding we mention that the
energy difference $E_{KS}[\rho^f]  - E_{KS}[\rho^i]$ can be written
with an expression which is formally linear in $\Delta D$ up to an
infinity order. van Leeuwen and Baerends showed  that it  be obtained as a path
integral along a path in the space of densities that
connects the initial and final densities.
A suitable path of the parameter $\alpha$,
from $\rho^i (\alpha=0)$ to $\rho^f (\alpha=1)$
with $\rho(\alpha)$  is a parametric function of $\alpha$ (e. g. $\rho^f(\alpha) = \rho^i + \alpha \Delta\rho$).
More precisely its explicit form is
\begin{equation}
\label{appeq:pathint}
        E_{KS}[\rho^f]  - E_{KS}[\rho^i] = \sum_{ij} \Delta D_{ij} \int_{0}^{1} F[\rho(\alpha)] d\alpha
\end{equation}
In practice the $\alpha$ integral can be done very accurately by some
Gauss numerical integration method  over the
[0,1] interval but it is rarely necessary to go beyond the mid point
or the Simpson rule. Noteworthy,
the expression of TS method
of Eq.~\ref{appeq:e2} can be obtained using mid point integral approximation
over the $\alpha$ parameter
in Eq.\ref{appeq:pathint} while
while its integration using Simpson rule gives
exactly the expression  derived by Ziegler and Rauk and shown
above in Eq.~\ref{appeq:corr3}.

\end{appendix}

\bibliographystyle{abbrvnat}
\bibliography{shortj,biblio}

\end{document}